\documentclass[12pt]{article}

\begin{document}

%%%%%%%%%%%%%%%%%%%%%%%%%%
% Vatche Sahakian's macros

\newcommand{\bb}{\begin{equation}}
\newcommand{\ee}{\end{equation}}
\newcommand{\bbb}{\begin{eqnarray}}
\newcommand{\eee}{\end{eqnarray}}
\newcommand{\diag}{\mbox{diag }}
\newcommand{\Str}{\mbox{STr }}
\newcommand{\Tr}{\mbox{Tr }}
\newcommand{\Det}{\mbox{Det }}
\newcommand{\C}[2]{{\lk [{#1},{#2}\re ]}}
\newcommand{\AC}[2]{{\lk \{{#1},{#2}\re \}}}
\newcommand{\kk}{\hspace{.5em}}
\newcommand{\vc}[1]{\mbox{$\vec{{\bf #1}}$}}
\newcommand{\mc}[1]{\mathcal{#1}}
\newcommand{\del}{\partial}
\newcommand{\lk}{\left}
\newcommand{\ave}[1]{\mbox{$\langle{#1}\rangle$}}
\newcommand{\re}{\right}
\newcommand{\pd}[1]{\frac{\del}{\del #1}}
\newcommand{\pdd}[2]{\frac{\del^2}{\del #1 \del #2}}
\newcommand{\Dd}[1]{\frac{d}{d #1}}
\newcommand{\sech}{\mbox{sech}}
\newcommand{\pref}[1]{(\ref{#1})}

\newcommand
{\sect}[1]{\vspace{20pt}{\LARGE}\noindent
{\bf #1:}}
\newcommand
{\subsect}[1]{\vspace{20pt}\hspace*{10pt}{\Large{$\bullet$}}\mbox{ }
{\bf #1}}
\newcommand
{\subsubsect}[1]{\hspace*{20pt}{\large{$\bullet$}}\mbox{ }
{\bf #1}}

\def\ie{{\it i.e.}}
\def\eg{{\it e.g.}}
\def\cf{{\it c.f.}}
\def\etal{{\it et.al.}}
\def\etc{{\it etc.}}

\def\e{{\mbox{{\bf e}}}}
\def\AA{{\cal A}}
\def\BB{{\cal B}}
\def\CC{{\cal C}}
\def\DD{{\cal D}}
\def\EE{{\cal E}}
\def\FF{{\cal F}}
\def\GG{{\cal G}}
\def\HH{{\cal H}}
\def\II{{\cal I}}
\def\JJ{{\cal J}}
\def\KK{{\cal K}}
\def\LL{{\cal L}}
\def\MM{{\cal M}}
\def\NN{{\cal N}}
\def\OO{{\cal O}}
\def\PP{{\cal P}}
\def\QQ{{\cal Q}}
\def\RR{{\cal R}}
\def\SS{{\cal S}}
\def\TT{{\cal T}}
\def\UU{{\cal U}}
\def\VV{{\cal V}}
\def\WW{{\cal W}}
\def\XX{{\cal X}}
\def\YY{{\cal Y}}
\def\ZZ{{\cal Z}}

\def\sinh{{\rm sinh}}
\def\cosh{{\rm cosh}}
\def\tanh{{\rm tanh}}
\def\sgn{{\rm sgn}}
\def\det{{\rm det}}
\def\trace{{\rm Tr}}
\def\exp{{\rm exp}}
\def\sh{{\rm sh}}
\def\ch{{\rm ch}}

\def\ell{{\it l}}
\def\str{{\it str}}
\def\lp{\ell_{{\rm pl}}}
\def\blp{\overline{\ell}_{{\rm pl}}}
\def\ls{\ell_{{\str}}}
\def\bls{{\bar\ell}_{{\str}}}
\def\bM{{\overline{\rm M}}}
\def\gs{g_\str}
\def\gym{{g_{Y}}}
\def\geff{g_{\rm eff}}
\def\eff{{\rm eff}}
\def\r11{R_{11}}
\def\kel{{2\kappa_{11}^2}}
\def\kten{{2\kappa_{10}^2}}
\def\lpten{{\lp^{(10)}}}
\def\alp{{\alpha '}}
\def\alpe{{{\alpha_e}}}
\def\le{{{l}_e}}
\def\aleff{{\alp_{eff}}}
\def\sqaleff{{\alp_{eff}^2}}
\def\tgs{{\tilde{g}_s}}
\def\talp{{{\tilde{\alpha}}'}}
\def\tlp{{\tilde{\ell}_{{\rm pl}}}}
\def\tr11{{\tilde{R}_{11}}}
\def\wtilde{\widetilde}
\def\what{\widehat}
\def\hlp{{\hat{\ell}_{{\rm pl}}}}
\def\hr11{{\hat{R}_{11}}}
\def\hf{{\textstyle\frac12}}
\def\coeff#1#2{{\textstyle{#1\over#2}}}
\def\CY{Calabi-Yau}
\def\lessapprox{\;{\buildrel{<}\over{\scriptstyle\sim}}\;}
\def\greaterapprox{\;{\buildrel{>}\over{\scriptstyle\sim}}\;}
\def\inbar{\,\vrule height1.5ex width.4pt depth0pt}
\def\IC{\relax\hbox{$\inbar\kern-.3em{\rm C}$}}
\def\IR{\relax{\rm I\kern-.18em R}}
\def\IP{\relax{\rm I\kern-.18em P}}
\def\Z{{\bf Z}}
\def\R{{\bf R}}
\def\One{{1\hskip -3pt {\rm l}}}
\def\sst{\scriptscriptstyle}
\def\osc{{\rm\sst osc}}
\def\lam{\lambda}
\def\lc{{\sst LC}}
\def\pr{{\sst \rm pr}}
\def\cl{{\sst \rm cl}}
\def\D{{\sst D}}
\def\bh{{\sst BH}}
\def\vev#1{\langle#1\rangle}

\newcommand{\Sym}{\mbox{{\bf Sym}}}
\newcommand{\Tless}{\mbox{{\bf Traceless}}}

\input{epsf}

\begin{titlepage}
\rightline{CLNS 01/1747}

\rightline{hep-th/0107180}

\vskip 2cm
\begin{center}
\Large{{\bf The large M limit of\\
Non-Commutative Open Strings\\
at strong coupling
}}
\end{center}

\vskip 1cm
\begin{center}
Vatche Sahakian\footnote{\texttt{vvs@mail.lns.cornell.edu}}
\end{center}
\vskip 12pt
\centerline{\sl Laboratory of Nuclear Studies}
\centerline{\sl Cornell University}
\centerline{\sl Ithaca, NY 14853, USA}

\vskip 2cm

\begin{abstract}
Two dimensional Non-Commutative Open String (NCOS)
theory, well-defined perturbatively, may also be studied
at strong coupling and for large D-string charge by making use of
the Holographic duality. We analyze the zero mode dynamics of a 
closed string in the appropriate background geometry and map the results onto
a sector of strongly coupled NCOS dynamics.
We find an elaborate classical picture
that shares qualitative similarities with the SL(2,R) WZW model.  In the
quantum problem, we compute propagators and part of the energy spectrum of the
theory; the latter involves interesting variations in the
density of states as a function of the level number, and energies
scaling inversely with the coupling.
Finally, the geometry exhibits a near horizon throat, associated with
NCOS dynamics, yet it is found that the whole space is available for
Holography. This provides a setting to extend the Maldacena duality
beyond the near horizon limit.

\end{abstract}

\end{titlepage}
\newpage
\setcounter{page}{1}

\section{Introduction and Summary}
\label{intro}

Within the realm of perturbation theory it was originally
formulated in, 
string theory has been known to entail unusual and sometimes remarkable
dynamics, such as 
non-local interactions and interesting spectra. Certain puzzles that had
plagued theoretical physics for many years were successfully 
addressed in this perturbative framework
due to such unique attributes of the theory. In more recent years,
sometimes through the use of supersymmetry, certain non-perturbative
aspects of the theory have been explored, revealing yet richer and
more remarkable dynamics.

An important new tool in this program is the
so-called Holographic duality~\cite{MALDA1}-\cite{ADSLECT}
that allows the controlled
study of the theory beyond a perturbative expansion. It is believed that
this principle is a general one; that it applies to the theory
expanded about many of its multitude of vacua.
In particular, our goal in this work is to focus on
Non-Commutative Open Strings (NCOS)~\cite{SWNC}-\cite{GMMS}
within Wound String theory~\cite{GOMISOOG,DGK1,DGK2}.
The latter corresponds to a certain sector of string
theories that is characterized by non-relativistic dispersion
relations. In this setting,
a particular
class of non-perturbative solitons, longitudinal D-strings bounded with 
fundamental string charge,
involve particularly interesting dynamics;
they can be described by a 
two dimensional, Lorentz invariant theory
of open strings, with the time and space coordinates non-commuting.
This is a useful setup as it focuses onto an important
attribute of string theories, non-locality, in a framework that is relatively
simple. And of most interest to us, through the use of the Holographic
duality, this non-local dynamics is computationally accessible 
at strong coupling. It is worthwhile emphasizing that this is a quite
unusual situation; whereby {\em stringy dynamics} can be explored
non-perturbatively.

The Holographic duality in this context
establishes a strong-weak coupling map between two string theories\footnote{
A system that is most similar to this in this regard is that of Little
String theory~\cite{LST1,LST2}; 
however, in that context, there has not been a good
understanding of what constitutes a perturbative expansion of the theory.}.
This map is not well understood, but may hold
important treasures for one's intuition.
There have been suggestions 
that this correspondence may be formulated directly on the world sheet level
~\cite{DGK2};
and indications of rich dynamics in the theory at strong
coupling~\cite{MALDARUSSO,HARMARKS,VVSNCOS,FASTER}. 
It even appears that the system is a prime setting to
understand Holography beyond the usual near-horizon scaling 
limit~\cite{DGK2}.

In conventional approaches to computations in holographic duals, 
one has a field theory on one side, and perturbative string theory
(in practice a supergravity)
on the other. The useful probes are often fields in the supergravity
propagating in some curved background 
geometry~\cite{WITHOLO,BALABULK1,BALABULK2}
(see however~\cite{WILSONMALDA}). 
In the NCOS analogue,
a central role must be played by probes of the dual geometry
that are strings. The novelty in the problem then is that
it involves studying the dynamics of closed strings in curved backgrounds,
and mapping results from this non-linear sigma model
onto physical quantities in NCOS theory.

In this work, we present an attempt in understanding strongly
coupled  dynamics in NCOS theory using this Holographic duality. 
We do this by studying, to a leading
order in a semi-classical approximation scheme, the dynamics of
an unexcited IIB
closed string in the background geometry cast about a bound state
of D strings and fundamental strings. This involves
the NCOS (or Wound String theory) scaling limit, and the background
space of interest becomes conformal to $AdS_3\times S^7$
(see also discussion in~\cite{BERMAN}), in the presence
of an NSNS B-field. Effects from the RR fields, and certain
corrections arising from the running dilaton, are found subleading
to the dynamics we consider. 
To be more concrete, our space is mapped by: seven coordinates on a seven
sphere; one radial coordinate we call $v$ (which is chosen dimensionless); 
and a time and a space coordinate
associated with the two dimensional NCOS theory, that we denote by
$t$ and $y$. We then have in the perturbative
NCOS setting the non-commutation relation
$[t,y]=i\theta$, with non-locality in time measured
by the scale $\theta$. The latter also sets the length scale 
in the NCOS Virasoro spectrum. Furthermore, we compactify the $y$ 
coordinate on a circle of circumference $\Sigma$.

\begin{figure}
\epsfysize=10cm \centerline{\leavevmode \epsfbox{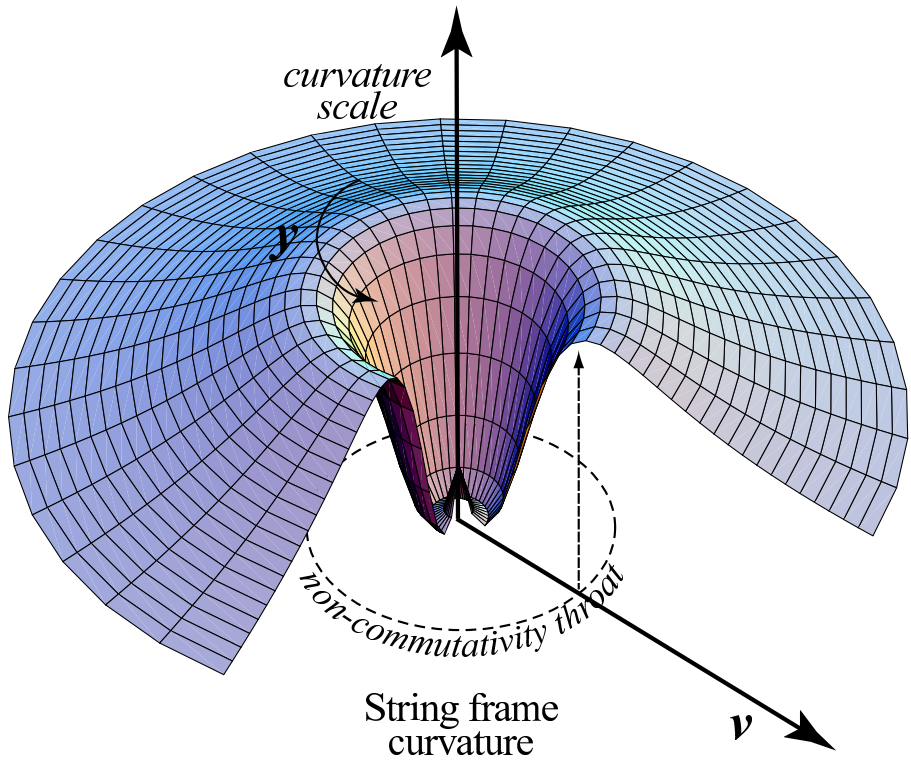}}
\caption{\sl
The string frame curvature (in IIB string
units) is finite everywhere, becoming zero at the center $v=0$ and 
at asymptotic infinity; and has a throat at $v^3\sim 1/G$. It is found that
the center $v=0$ repulses incoming strings.
}
\label{throat1}
\end{figure}

The $AdS_3$,
of curvature scale set by $\theta$, is then
parameterized by the coordinates $v$, $t$ and $y$. An important role
is however played by the conformal factor, which introduces 
additionally a throat in
the geometry at the 
radial coordinate $v^3\sim 1/G$; the dimensionless variable
$G$ being interpreted in the dual NCOS theory as the open string coupling. We
refer to this throat as the non-commutativity throat.
Figure~\ref{throat1} shows a depiction of the string frame scalar curvature,
as a function of $v$ and $y$. Note that this curvature is
bounded everywhere, being in particular zero at $v=0$ and $v\rightarrow\infty$. 
However, this does not mean that we can trust this background for all $v$;
as there are other conditions that need to be met to paint a reliable
picture of the dynamics.

Hence, the problem at hand is as follows. We consider the center of
mass motion of a closed string wrapping the cycle $y$ in the geometry
shown in Figure~\ref{throat1}, with zero angular momentum on the seven sphere. 
We analyze the motion first classically,
then quantum mechanically. This leads to several interesting
predictions about the dual NCOS theory; about a theory of strings
at strong coupling $G\gg 1$. In the subsequent paragraphs,
we summarize our main results.

\subsection{Classical dynamics}

The classical dynamics has similarities to a related system, 
the SL(2,R) WZW model~\cite{MALDAOOG1,MALDAOOG2,ADS2OOG}. 
The main differences arise from the presence of
the non-commutativity throat, which also makes the full problem in 
this case considerably more involved to solve for. The center of mass
dynamics of the closed string is however simple to unravel, as
it reduces to the dynamics of a point particle in a certain 
two dimensional curved space (the $v-t$ plane), in the presence
of a $v$ dependent potential and magnetic field.
We find an elaborate picture. 

\begin{figure}
\epsfysize=6cm \centerline{\leavevmode \epsfbox{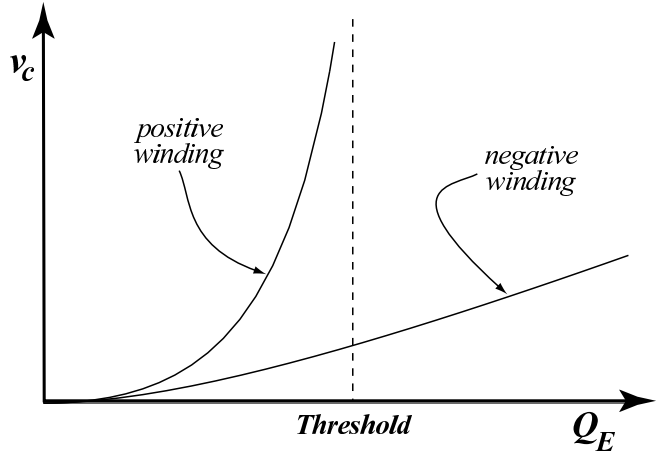}}
\caption{\sl
The maximum extent $v_c$ that the closed string reaches
in the radial direction $v$ as a function of its energy $Q_E$.
The `threshold' refers to the mass of the corresponding
wound string in the Wound string theory, $\omega \Sigma/(4\pi\alpe)$;
with $\omega$ being the winding number, and $\alpe$ the NCOS string scale
related to $\theta$ as in~\pref{thetaalpe}.
}
\label{vcfig}
\end{figure}

\begin{figure}
\epsfysize=5cm \centerline{\leavevmode \epsfbox{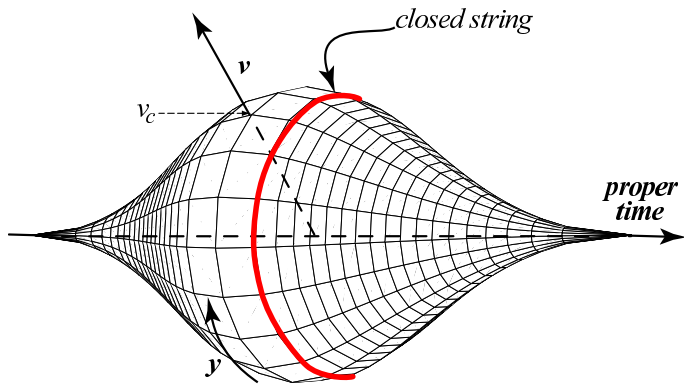}} 
\caption{\sl Depiction of the closed string world sheet
as a function of proper time for the bounded motions; both
possible winding orientations for the string generate qualitatively 
similar plots.
Note the deceleration effect near $v\sim 0$;
the closed string is repulsed from the center.
}
\label{wzw1}
\end{figure}

There are two sets of solutions that
describe bounded motion, differing by the orientation of the closed
string along $y$ (the analogues of `short strings' in the WZW model,
pointed out already in~\cite{KLEBMALDA,DGK2}). 
The string falls toward the horizon at $v=0$, initially
accelerating, then decelerating as it gets {\em repulsed} from
the origin. It comes to a halt at the origin and reverses direction,
but cannot escape to infinity. Instead, it reverses course again at some
finite $v=v_c$, and falls back in. The period of this
oscillatory motion is finite in proper time; but infinite as measured
by the time variable $t$, because of the infinite redshift associated with
the horizon. More interestingly, $v_c$ is not necessarily
within the non-commutativity throat! Depending on the classical 
energy of the closed string, $v_c$ can readily
extend to asymptotic infinity. Figure~\ref{vcfig} shows a plot of 
this maximum extent explored by the closed string
as a function of its energy $Q_E$. Positive (negative) 
winding refers to the fact that the
string is oriented parallel (anti-parallel) to the background B-field.
The throat region was
believed to be the analogue of the near horizon region of the Maldacena
duality~\cite{MALDA1,MALDA2}; 
perhaps being the `boundary' of the holographic projection. 
It was then suggestive that such bound state solutions would be confined
to within the throat area\footnote{This suggestion was made in~\cite{DGK2}
by analyzing the dynamics at small velocities.},
being duals to states in the NCOS theory.
We see however that this is not the case.
Figure~\ref{wzw1} shows a plot
of the dynamics; the two cases of different winding orientation
are qualitatively similar in this respect.

\begin{figure}
\epsfysize=5cm \centerline{\leavevmode \epsfbox{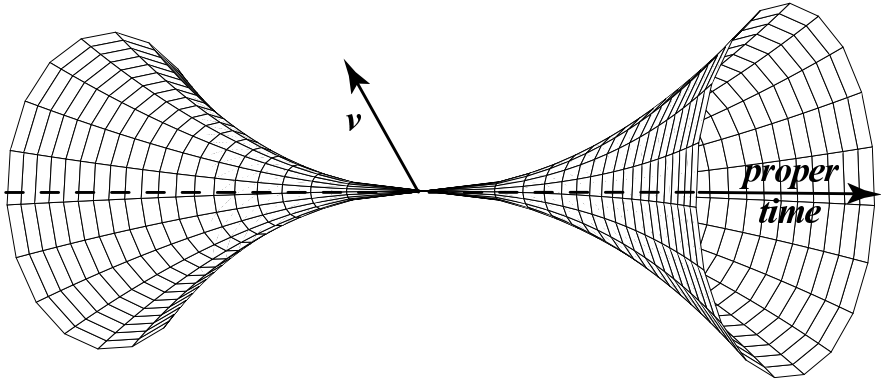}} 
\caption{\sl The closed string world sheet
as a function of proper time for the scattering scenario with
positive winding. 
}
\label{wzw2}
\end{figure}

If the energy of the string is high enough, only the positively wound closed
string can escape to infinity. The threshold energy is
given by the mass of the wound string in Wound String theory, which 
is consistent with one's intuition. We call the corresponding scenario
the scattering solution (the analogue of `long strings' in the SL(2,R)
WZW model). As shown in Figure~\ref{wzw2}, in finite proper time,
the closed string can be made to scatter off the horizon, if its
initial energy is high enough.
That the world-sheet theory at asymptotic
infinity coincides with that of the wound string was already shown 
in~\cite{DGK2}.
Hence, this presents a setting whereby one can define asymptotic 
on shell states corresponding to certain vertex operator insertions 
in the dual NCOS theory. This however also presents a puzzle, as the
time of scattering appears to be infinite in the variable $t$, due to the 
horizon redshift effect already encountered above.

This summarizes the classical dynamics of the problem. Next, we present
some of the results concerning the quantum mechanical treatment. 

\subsection{A spectrum}

\begin{figure}
\epsfysize=12cm \centerline{\leavevmode \epsfbox{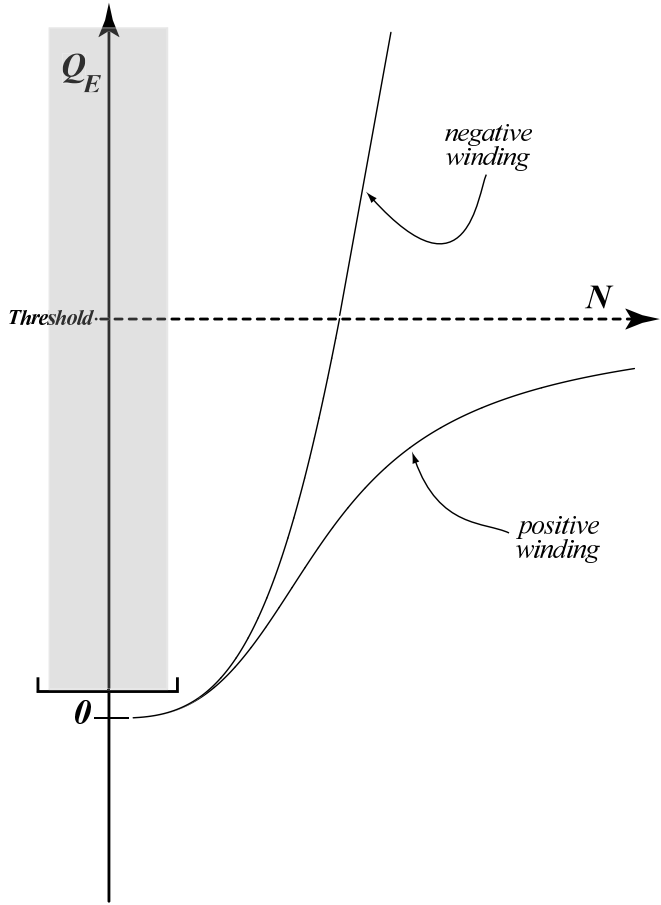}}
\caption{\sl The quantized spectrum associated with the bounded
dynamics of the closed string. $Q_E$ is mass of the state, and 
$N$ is an integer denoting the the level number.
This is a numerical plot of equations
~\pref{sclbnd}, \pref{sdW} and~\pref{born}. The threshold corresponds to
the rest mass of the corresponding wound string,
$Q_E=\omega \Sigma/(4\pi \alpe)$. The shaded area indicates
the window of energies for which the spectrum may be reliable,
as determined by the values of $M$ for the upper bound, and either
$\Sigma$ or $M$ for the lower bound (depending on whether $\sqrt{M}$ is smaller
or bigger than $\Sigma/\le$). As a general rule, 
the shaded area widens for large $M$ and large $\Sigma/\le$.
For this
plot, we have chosen
$G\sim 110$, $\Sigma\sim 23\le$, $\omega\sim 100$, and
the full scale on the $N$ axis is around $10^4$. The shaded area is drawn
for $M=1.5\times 10^5$; the upper bound is beyond the figure.
For $M\rightarrow\infty$, with all other parameters finite,
this shaded region encompasses the whole spectrum.
}
\label{specfig}
\end{figure}

Whenever one encounters bounded classical dynamics, one should expect
the possibility of a quantized energy spectrum in the quantum mechanical
problem. And indeed, due to subtle cancelations that appear
in the computation of the path integral of this problem, we argue that
the bounded dynamics leads to a prediction for part of the spectrum 
of the strongly coupled NCOS theory. Figure~\ref{specfig} summarizes these
results in a plot. For high energies,
the energy levels scale as $N^{3/2}/\sqrt{G}$ for
the negatively wound case, with $N$ being the level number, and $G$
the NCOS coupling; while they scale as $N^3/G$ for the positively wound
case for low energies
(the lower region in the figure, where the two spectra become
degenerate). The threshold energy indicated in the figure is the
mass of the corresponding wound closed string, which comes out correctly
from our analysis~\cite{KLEBMALDA}. And the shaded region
is to indicate that the computation can be trusted
for a window of energies; this is widened by increasing the number
of D-strings $M$, and/or increasing the cycle size $\Sigma$ in NCOS
string units. The analytical form for this spectrum is
presented in the text.

An interesting observation is that
all physical quantities we compute, classically or quantum mechanically, have
all instances of the variable counting number of longitudinal 
D-strings appearing only in a certain combination with the original
coupling of the theory. Denoting the open string coupling by $G_o$,
and by $M$ the number of D-strings, the effective coupling is argued to be
$G\equiv G_o\sqrt{M}$ for large $M$~\cite{VVSNCOS}. More importantly, the
strict limit $M\rightarrow \infty$ with fixed $G$ is {\em regular},
with all observables remaining finite and
with the background geometry becoming reliable for all $v$. In particular,
the center, generally associated with a singularity, is almost flat;
as if the non-commutativity throat plays the role of a regulator of
the singularity for $M\rightarrow\infty$. Consequently,
the spectrum is well-defined for all energies.
This constitutes one of the main results of this work;
a computation of part of
the non-perturbative spectrum of a string theory
at strong coupling and in the large $M$ limit.

At finite $M$,
the reliability of our result
hinges on certain assumptions about regularity
of the dynamics near $v\sim 0$, 
where string interactions become important. We
discuss this in some detail in the text; for large but finite $M$,
we have energy levels that may receive corrections
through an expansion in inverse powers of $M$.
We also argue for the possibility for
transitions between these various levels  with different
winding number sectors; the computed spectrum being however
still a good first order layout of the dynamics. Finally,
note also that we should superimpose, towering on each mass level
depicted in the figure, another spectrum corresponding to 
excitations of the closed string about the center of mass.

\subsection{Propagators}

In the NCOS theory at strong coupling, it then seems that one has
an interesting spectroscopy of states whose masses
scale inversely with the NCOS coupling. These objects perhaps may be
accorded a size, and corresponding breathing modes, that correlate
with location in the radial direction $v$ in the dual picture.
In the sector we confine our analysis, all these states carry zero
total momentum in the $y$ direction, as well as zero angular
momentum on the seven sphere. They can be attributed 
a `winding number' correlating with the number of times the correponding
closed string winds the $y$ direction in the dual description. 
We mentioned also scattering processes whereby a closed string in Wound
String theory, represented perhaps as a local vertex operator insertion
in the NCOS theory, scatters off the longitudinal D-strings.
All of this dynamics may be probed by computing propagators,
which we also do. For propagations of duration $T$ in 
NCOS string units, up to a `size'
(or location in the $v$ coordinate) $v_2$, we find that the propagators
diverge at
$T\sim 1/v_2$; and we find amplitudes that scale as $\sqrt{G}$ and $G$.

\subsection{Beyond the near horizon regime}

Another matter we focus on is an attempt to understand
in what sense Holography is being implemented in this setting; and,
in particular, to what extent is one studying this phenomenon
beyond the near-horizon limit that Holography is usually attributed with.
We observe that, while the throat is not a boundary of dynamics between
what we associate with NCOS theory, and what we qualify as Wound
closed string theory, it does however correspond to an energy scale at
which all running physical parameter of the theory scale as
those of the NCOS theory. The picture that seems to emerge is that
the whole of the space is holographically encoded in the NCOS theory,
but there is a screen at some definite location in the bulk,
where the non-commutative throat sits, that we may still
associate with a `projection' plane. While the Wound string theory
may be viewed as living at asymptotic infinity only. This strongly suggests
that there is a controlled approach
in which the Maldacena duality can be extended beyond the near horizon.

\subsection{Outline}

As this work was possibly more pleasant to type than it will be to read,
we have tried to organize the presentation such that
many computational matters are 
collected in appendices. Section 2 presents the notation and
the setup of the problem. Section 3 describes the classical dynamics.
Section 4 concentrates on the details of the quantum problem;
while Sections 5 and 6 collect the results and present some analysis.
Section 7 contains the observations with regards to the role
of the throat in the geometry. And Section 8 discusses various
loose ends, extensions and suggestions for future work.
The casual reader may focus on Sections 2, 3, 5 and 8 (and 
Appendix E if the need arises). Appendices A-E summarize certain
computational details.

\section{Preliminaries}
\label{prelim}

In this section, we review the two dimensional 
NCOS theory and establish notational matters. The geometry
dual to the strongly coupled NCOS theory is 
described first, with more background material on the subject collected
in Appendix A. We then outline the regime of validity of the setup.

\subsection{Background geometry}

Two dimensional NCOS theory describes the dynamics of a bound state
of D strings and fundamental strings in IIB string theory.
It can also be attained as
a subsector of Wound String theory in the presence
of longitudinal D-strings.

NCOS theory is parameterized by an integer $M$ that
counts the number of D-strings;
a string coupling $G$ restricted to a finite range
$0\leq G \leq M/(32\pi^2)$; a string length $\le$ that sets the
scale for (a) the spacing of the levels in the
Virasoro tower of free open string excitations; and (b) for a
parameter $\theta$ that measures non-commutativity of the coordinates
\bb\label{noncomm}
\lk[t,y\re]=i\ \theta\ .
\ee
Here, $t$ and $y$ denote the coordinates of this two dimensional 
theory.
In our conventions, we have
\bb\label{thetaalpe}
\theta= 2\pi \alpe\equiv 2\pi \le^2\ .
\ee
We will also choose to compactify the $y$ coordinate on a circle
of size $\Sigma$.

This theory has a well defined perturbative expansion
for $G\ll 1$ inherited from
the parent string theory. Novel features of the dynamics
can be attributed to the
non-commutation relation~\pref{noncomm}.

The strong coupling dynamics of the two dimensional NCOS theory can
be described via a dual setup, by studying IIB closed string theory
in the background geometry cast about a 
bound state of D-strings and fundamental
strings. A particular scaling limit is needed to establish this correspondence
and it is briefly outlined in
Appendix A. The resulting background is given by the metric (in the string
frame)
\bb\label{metric}
ds_{str}^2=\Omega^2 \lk\{ \frac{v^2}{8\pi^2 \alpe}
\lk(-dt^2+\Sigma^2 dy^2\re)
+\frac{dv^2}{v^2}
+4 d\Omega_7^2\re\}\ ,
\ee
where
\bb\label{omega}
\Omega^2\equiv 8\pi^2 \alp \frac{\sqrt{G}}{\sqrt{v}}
\sqrt{1+G v^3}\ ,
\ee
and $\alp$ is the string scale in the associated IIB theory.
The space is conformal to $AdS_3\times S^7$\ \footnote{
We work in the analogue of 
the Poincare patch of $AdS_3$~\cite{BALABULK1}; 
the
motivation for this is that, in the S-dual picture, the
corresponding geometry describes strongly coupled two dimensional
SU(N) gauge theory with an electric flux; with a Hamiltonian canonical to our 
time variable $t$.
}. As such,
dynamics in this background will have qualitative similarities
to that of the SL(2,R) WZW system. Note also that we have
rescaled the $y$ coordinate such that it is
compact of size $1$. Furthermore, our choice of coordinates
is such that $v$ is dual to energy in the NCOS theory in NCOS
string units (\ie\ $v$ is dimensionless); this is often
termed the UV-IR map in the Holographic duality~\cite{PEETPOLCH}.
The large $v$ region, corresponding
to the UV in the NCOS theory, is where, if the conformal factor
$\Omega$ was missing, the timelike $AdS_3$ boundary would be sitting.

The dilaton in this geometry runs as
\bb\label{dil}
e^\phi=\lk(32 \pi^2\re)^2 \frac{G^{3/2}}{M}
\frac{1+G v^3}{v^{3/2}}\ .
\ee
We also have an axion field
\bb
\chi=\frac{M}{\lk(32 \pi^2 G\re)^2} \frac{1}{1+G v^3}\ .
\ee
There is the RR two-form gauge field
\bb
A_{ty}=-\frac{\alp}{\alpe} \Sigma \frac{M}{\lk(32 \pi^2 G\re)^2}\ ;
\ee
and, most importantly, the NSNS B-field 
\bb\label{Bfield}
B_{ty}=\frac{\alp}{\alpe} \Sigma\ G v^3\ .
\ee

IIB string theory in this background is expected to be holographically
encoded into two dimensional NCOS theory.
We will focus on probing the dynamics in this geometry
using a closed string that wraps the cycle $y$. Figure~\ref{throat1}
shows a plot of the geometry at hand. The Penrose diagram is
identical to that of the Poincare patch of $AdS_3$ 
(see for example ~\cite{BALABULK1}).

\subsection{Regime of validity}
\label{valsec}

\begin{figure}
\epsfysize=7cm \centerline{\leavevmode \epsfbox{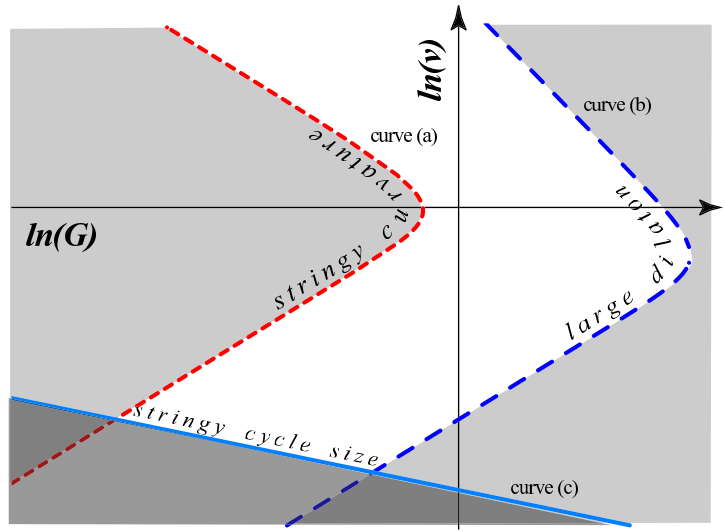}}
\caption{\sl The regime of validity of the background geometry used in
the text, as a function of the coupling $G$
and the energy scale (or coordinate) $v$.
The white area is where we confine our calculations.
To the left, perturbative NCOS dynamics can be used to study the system.
The corner in curve (a) is at $G\sim 1$; that
of curve (b) is at $G\sim \sqrt{M}$. They both occur along the
line $G v^3\sim 1$., \ie\ the non-commutativity throat of Figure~\ref{throat1}.
The right intersection point of curves (b) and (c) is around 
$G\sim \le \sqrt{M}/\Sigma$, which arises in our discussion on
several occasions. For a more complete analysis of this setup in a
thermodynamic setting, see~\cite{VVSNCOS}.
}
\label{validity}
\end{figure}

Given that
the geometry given by~\pref{metric} arises in a low energy limit of 
perturbative IIB
string theory, we need to determine the regime where this
setup is reliable. First,
the curvature scale must not be stringy,
which leads to the condition
\bb\label{curvcond}
\sqrt{\frac{v}{G}} \frac{-2+7 G v^3}{\lk(1+G v^3\re)^{3/2}}\ll\frac{1}{\alp}
\Rightarrow\lk\{
\begin{array}{ll}
v\gg 1/G & \mbox{  for  }G v^3\gg 1\mbox{  and  }G\ll 1 \\
v\ll G & \mbox{  for  }G v^3\ll 1\mbox{  and  }G\ll 1
\end{array}\re.\ ,
\ee
with no restriction for $G\gg 1$ (see the curve labeled (a)
in Figure~\ref{validity}).
The coordinate $v$ is dual to energy in the NCOS theory, 
in units of $\le$; hence, this is
a statement restricting energy scale. Requiring that the closed string
coupling be small yields the additional condition
\bb\label{curveb}
v\gg \frac{G}{M^{2/3}} \lk(1+G v^3\re)^{2/3}\Rightarrow
\lk\{
\begin{array}{ll}
v\ll M^{2/3} G^{-5/3} & \mbox{  for  }G v^3\gg 1 \\
v\gg M^{-2/3} G & \mbox{  for  }G v^3\ll 1
\end{array}\re.\ ,
\ee
or the curve labeled (b) in the figure below.
On the low energy (small $v$) side, the compact cycle $y$ becomes of 
stringy size unless 
\bb\label{curvec}
v^3\gg \frac{1}{G} \lk(\frac{\le}{\Sigma}\re)^{4}\ ,
\ee
\ie\ curve (c) in Figure~\ref{validity}. The region between these
three curves is the arena where we will confine our calculations. 

A key observation that we will rely on is that each of these three
curves is controlled my one of three independent parameters.
We first choose $G\gg 1$ so as to be safely onto the right
of curve (a). We then choose $M\gg 1$ to create a hierarchically 
wide regime of energies between the two flanks of
curve (b). And, finally, we make
$\Sigma/\le \gg 1$ so as to push curve (c) towards parametrically
smaller values of $v$.
The patch of spacetime then available for us extends from the
largest values
of $v$ to the smallest, a window controlled by $M$ and $\Sigma$.

Outside this domain, our calculations may be extended by applying
a sequence of dualities, as in~\cite{VVSNCOS}. However, for low or large enough
energies, and for finite $M$, 
we would eventually have to reach regions of space of
stringy curvature scales (or Planckian scales in an M theory), 
for both large (or small $v$). We will argue later
that, within the bounds of the space of parameters we will be working
in, our conclusions to leading order in an expansion in $1/M$
may be insensitive to the details behind
these stringy walls at large and small $v$. We will have to assume however
well-behaved boundary data in these asymptotic regions. 

More interestingly, 
we note that the only explicit dependence on $M$ appears in
the condition dictated by curve (b); and it is such that, for 
$M\rightarrow \infty$, with $G>1$ and all other parameters held finite
\footnote{Note that this also corresponds to the large $N$ limit
in the S-dual two dimensional SU(N) gauge theory (see equations~\pref{Gs}
and~\pref{cpl}).},
curve (b) imposes no restrictions on $v$. We are left with
curve (c) to worry about, which sets a lower bound on $v$. 
We then need to apply two duality transformations; first, a T-duality
on $y$, and then, for yet lower values of $v$, 
we need to lift to M-theory. Following~\cite{VVSNCOS}, we find
that the M-lift in the T-dual theory is needed at
\bb
v\sim \frac{G^{5/9}}{M^{4/9}} \lk(\frac{\le}{\Sigma}\re)^{4/9}\ .
\ee
But this goes to zero as $M\rightarrow\infty$, with all other parameters
held fixed. The is due to the fact that our metric and
B-field do not depend on $M$ explicitly, yet the dilaton scales
inversely with it. Hence, in the T-dual IIA theory, as in the original IIB
theory, the $M\rightarrow\infty$ limit renders the dilaton and string
interactions negligible. And the geometry 
can be trusted {\em for all $v$}, as there are no other restrictions
that arise (see~\cite{VVSNCOS} for the details). The physical conclusions
will be unchanged under the application of the duality
transformation, as this corresponds to changing framework
within the same parent theory; describing the same dynamics with
other degrees of freedom. And
all physical observables we compute will be found independent of $M$.
This strict limit is then indeed regular, and
our computations are reliable when restricted to this
regime. This also gives us partial
confidence that the finite $M$ case, for fixed but large
values of $M$, gives a faithful picture of the dynamics to leading
order in an expansion involving inverse powers of $M$.

We also note that
a number of papers have studied dynamics of the theory onto the left
of Figure~\ref{validity}, 
where a perturbative expansion in $G$ can be trusted~\cite{GGKRW,KRISRAJAN}.

\section{Classical dynamics}

In this section, we probe the geometry given by~\pref{metric} by
studying the dynamics of a closed string wrapping the cycle $y$.
In the first subsection, we set up the action, ansatz and 
classical equations of motion. In the second
subsection, we focus on the specific case of interest, solve the 
corresponding
dynamics, and analyze the resulting classical motion of the wrapped
string.

\subsection{Equations and ansatz}

The classical motion of a closed string in a curved background is
described by the action (see for example~\cite{POLCHV1})
\bb\label{Saction}
S=\frac{1}{4\pi \alp} \int d^2\sigma\ 
\sqrt{-h} \lk\{
\lk( -h^{ab} G_{\mu\nu}(X)+\varepsilon^{ab} B_{\mu\nu}(X)\re)
\del_a X^\mu \del_b X^\nu
+\LL_{dil}
+\LL_{ferm} \re\}\ ,
\ee
where $h_{ab}$ is the worldsheet metric. The coupling to the
dilaton and RR fields is given by
\bb
\LL_{dil}= \alp R^{(2)} \phi(X)\ ,\ \ 
\LL_{ferm}=\LL_{free}+\LL_{RR}\ ,
\ee
where $\LL_{free}$ contains the kinetic term for the fermions
(and the kappa symmetry term in the spacetime formalism), while
$\LL_{RR}$ contains the coupling of the worldsheet theory
to the background RR fields through fermions\footnote{
The latter would involve terms of the form
$\del_\mu \chi \bar{\VV} C \Gamma^\mu \VV$ and
$\del_{[\alpha} A_{\beta\gamma]} \bar{\VV} 
C\Gamma^{\alpha\beta\gamma} \VV$ that couple the spin fields
to the axion and
D-string gauge field. In this respect,
it is wiser to formulate such a setup from the outset
with the spacetime supersymmetry formalism, since
the vertex operators coupling to the RR fields are then more transparent
(see also~\cite{BVW}).}.
The details of the fermionic contributions will be irrelevant, as
we will focus in this work
on classical bosonic dynamics on the worldsheet, 
with the fermion fields set to zero\footnote{One effect at this level however
may be additional degeneracies in the spectrum
from fermionic zero modes.} . Furthermore, the coupling
to the dilaton is subleading, being weighed
by a power of the string length $\alp$. The dynamics we will find can be 
plugged back into our energy-momentum tensor, 
to analyze the justification of dropping
the term given by $\LL_{dil}$. One then finds the condition~\pref{curvcond}.
Hence, for the setup we will be concerned with, valid within the region
depicted in Figure~\ref{validity}, 
both $\LL_{dil}$ and $\LL_{ferm}$ may be ignored.

The worldsheet theory is conformal by definition. The metric $h_{ab}$
can be fixed, while supplementing the equations of motion with the
first class constraint
\bb\label{constraint}
G_{\mu\nu} \del_c X^\mu \del_d X^\nu
-\frac{1}{2} h_{cd} G_{\mu\nu} h^{ab} \del_a X^\mu \del_b X^\nu
+O(\alp)=0\ ;
\ee
otherwise known as the string on-shell condition. Note that we have 
dropped from these equations contributions from the worldsheet 
quantum field theory of order $\alp$ and beyond\footnote{Such terms 
typically appear as quantum corrections to 
all components of the worldsheet energy-momentum tensor; in
addition to correcting its trace, where they add to
the equations of 
motion of IIB supergravity, as the condition for scale invariance on
the worldsheet.}. Also, there is no contribution from the background
$B$ field.

The equations of motion that follow from~\pref{Saction} are
\bb\label{eom}
\nabla_a \nabla^a X^\gamma=
-\Gamma^\gamma_{\mu\nu} h^{ab} \del_a X^\mu \del_b X^\nu
+\frac{1}{2} G^{\gamma\alpha} \varepsilon^{ab} H_{\alpha \mu\nu}
\del_a X^\mu \del_b X^\nu\ ,
\ee
with $\Gamma^\gamma_{\mu\nu}$ the Christoffel variables associated with
the background metric $G_{\mu\nu}$\footnote{Throughout, we conform
to the conventions in~\cite{WALD}.}. And the field strength $H$ is defined as
\bb
H_{\alpha\mu\nu}\equiv 3 \del_{[\alpha} B_{\mu\nu]}\ .
\ee

With the choice of coordinates given by the background~\pref{metric},
we write the worldsheet scalars $X^\mu$ as
\bb
X^\mu\in\{t(\tau,\sigma),v(\tau,\sigma),y(\tau,\sigma),\Theta(\tau,\sigma)\}\ ,
\ee
where $\Theta$ denotes all angle variables on the seven sphere
of~\pref{metric}. We also fix the worldsheet metric to $h_{ab}=\eta_{ab}$.

The classical system described by equation~\pref{eom}
can be solved by~\cite{BOZHILOV,MATTOS}
\bb
X^\mu(\sigma+\tau)\mbox{  or  }X^\mu(\sigma-\tau)\ ,
\ee
for otherwise arbitrary functions,
irrespective of the background fields. Generally, we loose however
the naive principle of superposition. In special backgrounds
one may find solutions with simultaneously
both left and right moving modes.

We focus instead on dynamics subject to the ansatz
\bb
t(\tau,\sigma)\rightarrow t(\tau)\ ,\ \ 
v(\tau,\sigma)\rightarrow v(\tau)\ ,\ \ 
y(\tau,\sigma)\rightarrow \pm \omega \frac{\sigma}{\Sigma}\ ,\ \ 
\Theta(\tau,\sigma)\rightarrow 0\ .
\ee
$\omega$ is the number of windings of the closed string on $y$
($\sigma$ has size $\Sigma$). We have chosen 
\bb
\omega \geq 1\ ,
\ee
and will account for both orientations explicitly by the $\pm$ sign
in subsequent equations. The upper choice will be referred to as positive
winding, the lower as negative. Positive winding corresponds to the
closed string oriented parallel to the background B field.

Our ansatz corresponds to the center of mass motion of the closed
string. Furthermore, if we were to consider the slight generalization
\bb
y(\tau,\sigma)\rightarrow \pm \omega \frac{\sigma}{\Sigma}+y(\tau)\ ,
\ee
it is easy to check that the constraint from the off-diagonal elements
of the worldsheet energy-momentum tensor requires $y(\tau)$ to be constant.
Roughly speaking, we cannot put momentum on the probe string without 
wiggling it.

Subject to this ansatz, and in the background geometry 
given by~\pref{metric}-\pref{Bfield}, the Lagrangian becomes 
\bb\label{lagrangian}
\LL=\pm \omega \frac{\Sigma}{2\pi \alpe} G v^3 \dot{t}
+\frac{\Sigma}{4\pi} \frac{\Omega^2}{\alp} \frac{{\dot{v}}^2}{v^2}
-\frac{\Sigma}{32 \pi^3 \alpe} \frac{\Omega^2}{\alp} v^2 {\dot{t}}^2
-\frac{\Sigma}{32 \pi^3 \alpe} \frac{\Omega^2}{\alp} v^2 \omega^2\ ;
\ee
\ie\ a classical system describing a point particle, 
in the presence of a background gauge field and a potential,
evolving in proper time $\tau$ (see~\cite{BOZHILOV} 
for a more general treatment); 
we denote
\bb
\dot{v}\leftrightarrow \frac{dv}{d\tau}\ ,\ \ 
\dot{t}\leftrightarrow \frac{dt}{d\tau}\ .
\ee
And we leave the $\Omega$ factor arbitrary in this subsection.

The Hamiltonian becomes
\bb\label{hamil}
H=\frac{\pi}{\Sigma} \frac{\alp}{\Omega^2} v^2 \Pi_v^2
-\frac{8\pi^3 \alpe}{\Sigma v^2} \frac{\alp}{\Omega^2}
\lk(\Pi_t\mp\omega \frac{\Sigma}{2\pi\alpe} G v^3\re)^2
+\omega^2 \frac{\Sigma}{32 \pi^3 \alpe} \frac{\Omega^2}{\alp} v^2\ ,
\ee
with the canonical momenta given by
\bb\label{Piv}
\Pi_v=\frac{\Sigma}{2\pi} \frac{\Omega^2}{\alp} \frac{\dot{v}}{v^2}\ ;
\ee
\bb\label{Pit}
\Pi_t=\pm\frac{\Sigma}{2\pi \alpe} v^2 
\lk(G\omega v\mp\frac{\Omega^2}{8\pi^2 \alp} \dot{t} \re)\ .
\ee

The equation of motion for the radial coordinate $v$ is then
\bb\label{radialeom}
2\frac{\Omega^2}{v^2} \ddot{v}
-2\frac{\Omega^2}{v^2} {\dot{v}}^2
\lk(\frac{1}{v}-\frac{\Omega'}{\Omega}\re)
+\frac{v^2\Omega^2}{4\pi^2 \alpe} 
\lk(\frac{1}{v}+\frac{\Omega'}{\Omega}\re) {\dot{t}}^2
\mp 6\omega G \frac{\alp}{\alpe} v^2 \dot{t}
+\omega^2 \frac{v^2\Omega^2}{4\pi^2 \alpe}
\lk(\frac{1}{v}+\frac{\Omega'}{\Omega}\re)=0\ ;
\ee
while for the time coordinate $t$, it is 
\bb\label{timeeom}
\dot{\Pi}_t=0\ ;
\ee
\ie\ $\Pi_t$ is the N\"{o}ether charge associated with the Killing
vector field $\del_t$.

The constraint equation becomes
\bb\label{expconst}
\frac{1}{2} \frac{\Omega^2}{v^2} {\dot{v}}^2
-\frac{v^2 \Omega^2}{16\pi^2 \alpe} {\dot{t}}^2
+\omega^2\frac{v^2\Omega^2}{16\pi^2\alpe}=0\ ,
\ee
\ie\ the vanishing of the worldsheet Hamiltonian $H=0$,
which can also be interpreted as the on-shell condition for a particle in
a certain curved background. As the system is
constrained, one of the two solutions of
~\pref{radialeom} must be dropped in view of~\pref{expconst}. 
Of course, consistent
dynamics requires that the constraint evolves properly by the equations
of motion, as it does. 

We solve the $t$ equation~\pref{timeeom} trivially by introducing a constant of
motion $E$
\bb\label{Edef}
\Pi_t=\frac{\Sigma}{4\pi\alpe} E\ .
\ee
$E$ is a dimensionless parameter related to the energy
of the probe $Q_E$ as seen by the dual NCOS theory
(canonical to $\delta t$) as in
\bb\label{Qdef}
Q_E\equiv -\Pi_t\ .
\ee
The factor $\Sigma$ introduced in~\pref{Edef} helps to make our intermediate
equations somewhat less cluttered. Note the negative sign
in~\pref{Qdef}, which we justify below.
Using~\pref{Edef} and~\pref{Pit}, we then find
\bb\label{tdot}
\dot{t}=\frac{4\pi^2\alp}{\Omega^2}
\frac{-E\pm 2 G \omega v^3}{v^2}\ .
\ee
This allows us to solve the constraint equation~\pref{expconst} for $\dot{v}$
\footnote{In taking the square root of~\pref{vdot}, we choose conventionally
$\dot{v}<0$ throughout, 
\ie\ in-bound motion. The reader may assume that
in all subsequent equations where this ambiguity arises this choice
has been made.}
\bb\label{vdot}
{\dot{v}}^2 =\frac{2\pi^2 \alp^2}{\alpe\Omega^4}
\lk(\lk(E\mp2 G \omega v^3\re)^2
-\omega^2 \frac{v^4\Omega^4}{16\pi^4 \alp^2}\re)\ .
\ee
The two possible signs for $\dot{v}$ 
correspond to motion inward and outward from 
the center of the geometry located at $v=0$. It is easy
to check that this solution satisfies~\pref{radialeom}. 

One way to think of the situation is in analogy to Light-Cone gauge
fixing in flat backgrounds. We would like to gauge fix say some $X^+\sim \tau$
using the residual conformal symmetry on the worldsheet; residual 
after fixing $h_{ab}=\eta_{ab}$. We could do this in flat space since 
this choice would solve the equations of motion. In our curved background,
the analogous fixing of the residual conformal symmetry, which we still
have, is solving for the time coordinate $t$ using the time translational
invariance of the background. This is given implicitly by~\pref{tdot}.
Once equation~\pref{vdot} is integrated, we obtain
$t(\tau)$ from~\pref{tdot}. And~\pref{expconst} is nothing but the constraint
that in the flat space case allows us to solve for $X^-$.
These basic ideas are illustrated also when setting up the Fateev-Popov
determinant in the path integral in Appendix B. Finally, 
note that, because the system is constrained, we loose one constant of motion.
The dynamics is parameterized by $E$, and the initial and final positions of
the variable $v$; we denote them by $v_1$ and $v_2$, arising
as integration limits in~\pref{tdot} and~\pref{vdot}.

\subsection{Analysis of the solutions}

A more instructive way to view the dynamics is to push the
analogy with a point particle further. We are solving for
\bb
\frac{{\dot{v}}^2}{2}+V(v)=\EE=0\ ,
\ee
with
\bb\label{pot}
V(v)\equiv -\frac{\pi^2 \alp^2}{\alpe \Omega^4}
\lk( \lk(E\mp 2 G\omega v^3\re)^2
-\omega^2 \frac{v^4\Omega^4}{16\pi^4 \alp^2}\re)\ .
\ee
By studying the shape of this potential $V(v)$, we see the bulk
properties of the motion. At this point, we will need to
use the explicit form of $\Omega$ given by~\pref{omega}.

\begin{figure}
\epsfysize=6cm \centerline{\leavevmode \epsfbox{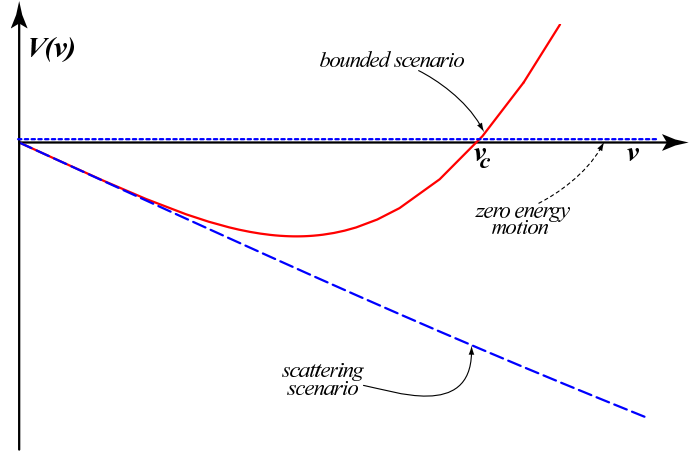}}
\caption{\sl The potential given by equation~\pref{pot}
as a function of the radial coordinate $v$.
The `particle' has zero total (worldsheet) energy, corresponding to the
on-shell constraint condition on the dynamics. 
Two solutions are associated with the bounded scenario, with 
both orientations for closed string winding; and one solution with
positive winding corresponds to a scattering process. 
}
\label{potfig}
\end{figure}

From equations~\pref{pot} and~\pref{omega}, we see that
\bb
V(v)\rightarrow \frac{E^2}{64\pi^2 \alpe G^2} \frac{v}{v_c^3}
\mbox{  as  }v\rightarrow \infty\ ,
\ee
due to a subtle cancelation of terms, irrespective of the
winding orientation. We have introduced the parameter
\bb\label{vc}
v_c^3\equiv\frac{E^2}{4 G \lk(\omega^2\pm \omega E\re)}\ ,
\ee
whose significance will become apparent below. 
In the opposite limit, we have
\bb
V(v)\rightarrow -\frac{E^2}{64\pi^2 \alpe G} v
\mbox{  as  }v\rightarrow 0\ .
\ee
for both orientations.
And finally
\bb
V(v)=0\Rightarrow v=v_c\mbox{  or  }v=0\ .
\ee
It is now easy to plot the potential $V(v)$. There are two qualitatively
different cases. For $v_c^3>0$, we have bounded motion with
$0\leq v \leq v_c$. While for $v_c<0$, 
we have unbounded motion $0\leq v$ (see Figure~\ref{potfig}).

Using~\pref{omega}, one easily gets the explicit form of the evolution of the
variables
\bb\label{tdot2}
\dot{t}=\frac{-E\pm2 G \omega v^3}
{2\sqrt{G} v^{3/2} \sqrt{1+G v^3}}\ .
\ee
\bb\label{vdot2}
{\dot{v}}^2=\frac{E^2}{32\pi^2 \alpe G}
\frac{1-(v/v_c)^3}{1+G v^3} v\ .
\ee
One then integrates~\pref{vdot2}, and substitutes in~\pref{tdot2}.
Note that $\dot{v}$ vanishes for all $v$ for $E=0$.
 
We next determine the physical relevance of the various possible
solutions. We need to require that
\bb\label{ftime}
\frac{dt}{d\tau}\geq 0\Rightarrow
-E\geq \mp 2 G \omega v^3\ .
\ee
First, consider the case $v_c^3<0$, \ie\ unbounded motion. It is then easy
to see that the only possibility consistent with~\pref{ftime} is
for the winding of the string to be parallel to the B-field; negative
winding is not possible, as it corresponds to opposite propagation 
in times $t$ and $\tau$. Furthermore, the combination of the two conditions 
$v_c^3<0$ and~\pref{ftime} in the
case of positive winding leads to $E<-\omega$.

Next consider the case $v_c^3>0$, \ie\ bounded motion.
We then find that the positive winding solution must satisfy
$-\omega<E<0$, complementing the parameter space established 
by the unbounded case. However, looking at the negative winding case,
we now find that it is also a possible solution to the dynamics. 
For this scenario, assuming $E>0$ leads to
\bb
E^2\geq 2 E \omega\mbox{ and } E\leq \omega\ ;
\ee
a contradiction. While $E<0$ leads to the consistent scenario.

To validate the proposed dynamics, we also look at the `acceleration'
at two critical points in the motion. We find
\bb\label{vdd}
\lk.\ddot{v}\re|_{v=0}=\frac{E^2}{64\pi^2 \alpe G}\geq 0\ ;
\ee
and 
\bb\label{vdd2}
\lk.\ddot{v}\re|_{v=v_c}=-\frac{3 E^2}
{64\pi^2\alpe G \lk(1+G v_c^3\re)}\leq 0\ .
\ee
From~\pref{vdd}, we see that the string is repulsed from the origin
$v\sim 0$, where a horizon sits.

We can now put the story together, depicting it as in Figures~\ref{wzw1}
and~\ref{wzw2},
in analogy to classical dynamics that has risen in the context of 
the SL(2,R) WZW model. We have two oscillating solutions which we call
bound states, and one unbounded motion that reaches asymptotically
larger values of $v$; we call the 
latter a scattering solution. We summarize the scenarios as follows
\bb\label{table}
\lk\{
\begin{array}{lclcc}
v_c^3<0 & \Rightarrow & \mbox{ unbounded dynamics with positive winding } 
& \Rightarrow & E<-\omega \\
v_c^3>0 & \Rightarrow & \mbox{ bounded dynamics with positive winding } 
& \Rightarrow & -\omega<E<0 \\
v_c^3>0 & \Rightarrow & \mbox{ bounded dynamics with negative winding } 
& \Rightarrow & E<0 \\
\end{array}
\re.
\ee

We note that for all cases, we have
\bb
E< 0\ .
\ee
However, in our chosen convention for $E$ through~\pref{Edef}, there is
a sign difference between real energy $Q_E$
and $E$ as seen in~\pref{Qdef}. Hence,
the N\"{o}ether energy $Q_E$ is positive definite for all cases.
The origin of the sign flip will be seen below. It is also significant that the
threshold energy $E=-\omega$ is then identified in the dual picture
as $Q_E=\omega \Sigma/(4\pi\alpe)$, the mass of the closed string in
Wound String theory, with the additional $1/2$ factor for
the wound string tension that has already been noted in~\cite{KLEBMALDA}. 
This indicates
that we have identified in this dual picture what corresponds to energy
in NCOS theory with the correct numerical factor.

The two solutions with positive winding have an attractive interpretation
in the dual NCOS theory. We will argue below that the scattering
solution is seen as a closed string in Wound string theory 
scattering off the bound state of D strings and fundamental strings.
In the NCOS theory, it is seen as an insertion of 
vertex operators, and the propagation of a non-perturbative open string
resonance in between\footnote{We note that in 
the embedding Wound string theory,
we are not working in a decoupling limit; our discussion naturally maps to
the analogue of extending the Maldacena duality beyond the near horizon
region (see discussion along this line of thought in~\cite{DGK2}.)}.
On the other hand,
if the corresponding open string state does
not have enough energy to create the appropriate closed string, as in 
the case of the bounded solution with positive winding and $-\omega<E<0$,
it is a non-perturbative state within NCOS theory.

The negatively wound
solution however is at first sight
a pathological case. It appears to 
correspond to a non-perturbative state in NCOS theory that cannot
leave the system no matter how high an energy in attains.
For now, we want to check the relevance of these
solutions quantum mechanically, with the hope that perhaps the negatively wound
state becomes metastable.

\section{Quantum effects}
\label{quantumeff}

The task in this section is to understand the quantum mechanics of the
classical dynamics we discussed above . It is a problem
of quantum mechanics as we have reduced the system to that of a point
particle moving in a curved space subject to an electric field and
a constraint. Such a treatment however assumes that quantum fluctuations
that can excite modes on the probe closed string outside
our ansatz are negligible. 
The quantum mechanics  we study
is that of the center of mass in a two dimensional
cross-section of the whole spacetime.
We will address the relevance of any additional corrections
in the Discussion section.

The propagator for the system is given by the path integral\footnote{
We normalized the asymptotic states as in
\bb
\langle v_2, t |v_1, t \rangle=\delta(v_2-v_1)\ ,
\ee
without the conventional $1/\sqrt{|g(v_1,t)|}$ factor on the right ($g$ being
the determinant of the metric). This is so that we may directly look at 
$|\GG|^2$ as probability of propagation, without the need to 
multiply with the measure in the curved space.
}
\bb\label{prop}
\langle v_2, \Delta t |v_1,0 \rangle\equiv
\GG\lk(v_1,v_2,\Delta t\re)\sim \int\ \DD v \DD \Pi_v
e^{i\int_1^2\ dt\ 
\lk(
\Pi_v \frac{dv}{dt} +\Phi_t\re)}\ ,
\ee
where
\bb\label{Phit}
\Phi_t=
\frac{v^2}{\sqrt{8\pi^2 \alpe}}
\sqrt{\Pi_v^2+\omega^2 \frac{\Sigma^2}{32\pi^4 \alpe} 
\frac{\Omega^4}{\alp^2}}\pm\omega \frac{\Sigma}{2\pi\alpe} G v^3\ .
\ee
The latter is just the constraint solved for $\Phi_t\rightarrow \Pi_t$.
This expression is such that the propagator is to be viewed as
a function of the initial and final positions in $v$, that we call
$v_1$ and $v_2$, and the
time separation $\Delta t$. It would tell us the probability
to propagate from $v_1$ to $v_2$ in time $\Delta t$.
Note that we were careful in writing this propagator in the Hamiltonian
formalism, as we are dealing with a path integral in curved space. 
The derivation of~\pref{prop} is given in detail in Appendix B. 
As seen from equation~\pref{prop}, the
system effectively evolves in time $t$ according to $-\Phi_t$.
This is the origin of the minus sign appearing in~\pref{Qdef}.

The result of this path integration is, to order $\hbar^2$
\bb\label{propres}
\GG\lk(v_1,v_2,\Delta t\re)\sim 
\sqrt{|\Delta|} e^{i S_{cl}+iS_{dW}}\ .
\ee
$S_{cl}$ is the action evaluated at the classical solutions given
by~\pref{tdot2} and~\pref{vdot2},
with the appropriate boundary conditions; $|\Delta|$
is determinant of the propagation operator in this curved space; and $S_{dW}$
is a term proposed by DeWitt~\cite{DEWITT} and is generally needed for
path integrals in curved spaces. Corrections to~\pref{propres} start at
order $\hbar^2$.  
In the subsequent subsections, we elaborate on these
three quantities and compute them.

\subsection{Leading contribution}

The leading contribution is obtained by substituting in
the exponent of~\pref{prop}
the classical action evaluated at the solutions given by~\pref{tdot2} and
~\pref{vdot2}. This is because the extremum of the exponential is at
these solutions for given boundary conditions, as shown in
Appendix B. Changing integration variable to $v$, we then get from
~\pref{lagrangian}, \pref{tdot2} and~\pref{vdot2}
\bb\label{sclint}
S_{cl}=\pm \omega G \Sigma \sqrt{\frac{2}{\alpe}} 
\frac{E\pm 2\omega}{-E}
\int_{v_1}^{v_2}\ dv\ 
\frac{v}{\sqrt{1-(v/v_c)^3}}\ .
\ee
The dependence of this expression on $\Delta t$ is implicitly
through the variable $E$, obtained by integrating and inverting
\bb\label{tint}
\Delta t=\sqrt{8\pi^2} \frac{\le}{E}
\int_{v_1}^{v_2}\ dv\ 
\frac{-E\pm2 \omega G v^3}{v^2\sqrt{1-(v/v_c)^3}}\ ;
\ee
the latter is simply~\pref{tdot2} divided by the square root of~\pref{vdot2}.
From~\pref{vdot2}, we also have an expression for the amount of proper
time it takes for the string to propagate between $v_1$ and $v_2$
\bb\label{tauint}
\Delta\tau=\sqrt{32\pi^2} \frac{\sqrt{G\alpe}}{E}
\int_{v_1}^{v_2}\ dv\ 
\frac{\sqrt{1+G v^3}}{\sqrt{v} \sqrt{1-(v/v_c)^3}}\ .
\ee

All these expressions can be solved in terms of various hypergeometric
functions. The more interesting physics is however in their asymptotic
forms; which we now write down.

We will distinguish two processes: a positive winding scattering
scenario; and two bound state scenarios with both possible winding
orientations. For the scattering scenario, we take $v_1$ very large,
and $v_2$ very small. For the bound state cases, we take $v_1=v_c$
and $v_2$ small. 
Making use of the appropriate boundary conditions in $v$ in each case,
as well as on $E$ as given by~\pref{table}, we find:

\begin{itemize}

\item For the bound states, we have
\bb\label{sclbnd}
S_{cl}^{bnd}=-\frac{\sqrt{2\pi}}{3} \frac{\Gamma(2/3)}{\Gamma(7/6)}
\frac{G}{|E|}
\frac{\Sigma}{\le}
v_c^2 \lk(\mp\omega |E|+2 \omega^2\re)\ ,
\ee
which is finite for both orientations of winding.

\item The action for the scattering process becomes
\bb\label{sclscatt}
S_{cl}^{scatt}\simeq
\frac{\Sigma}{\le} 
\sqrt{2 G \omega} 
\frac{|E|-2\omega}{\sqrt{|E|-\omega}} \sqrt{v_1}
-\frac{\Sigma}{\sqrt{2\alpe}} G \omega
\frac{|E|-2\omega}{|E|} v_2^2\ ,
\ee
and is large for $v_1$ big.

\item The time $\Delta t_{bnd}$
for half the trip in the bound state scenario is given by
\bb\label{deltatbnd}
2\Delta t_{{bnd}} \simeq \frac{\sqrt{32\pi^2 \alpe}}{v_2}\ ,
\ee
for both winding orientations.
The divergence, as $v_2\rightarrow 0$, is essentially
the infinite redshift that an external
observer generically witnesses as a probe approaches
a horizon. It
will play a crucial role in regulating the bound state spectrum later.

\item The scattering time in the same coordinate is given by
\bb\label{deltatscatt}
2\Delta t_{scatt}\simeq
\frac{\sqrt{128\pi^2 G \omega \alpe}}{\sqrt{|E|}} \sqrt{v_1}
+\frac{\sqrt{32\pi^2 \alpe}}{v_2}\ ,
\ee
which is large for both big $v_1$ and small $v_2$.

\item The half period in proper time $\Delta \tau_{bnd}$ that the string
takes to oscillate in the bound state scenario is
\bb\label{deltataubnd}
2\Delta \tau_{bnd}=
8\pi^2 \frac{\sqrt{2 G\alpe}}{3 |E|} 
\sqrt{v_c} \sqrt{1+G v_c^3}\ ,
\ee
which is finite unlike~\pref{deltatbnd}.

\item The proper time elapsed during the scattering process is
\bb\label{deltatauscatt}
2\Delta\tau_{scatt}\simeq
16\pi \frac{\sqrt{2 \alpe}}{E}
G \sqrt{|v_c|^{3}} \sqrt{v_1}\ ,
\ee
and increases unboundedly with large $v_1$.

\end{itemize}

We will use all these ingredients in upcoming sections
in attempting to understand 
strongly coupled dynamics in the dual NCOS theory.  Let us first
however 
compute the subleading terms to the propagator. 

\subsection{Sub-leading contributions}

\subsubsection{The determinant}

The determinant $\Delta$ appearing in~\pref{propres} is nothing more that 
\bb\label{unitarity}
|\Delta|=\lk|\frac{\del^2 S_{cl}}{\del v_1 \del v_2}\re|\ .
\ee
This is a well known statement from quantum mechanics, yet not
properly advertised or emphasized in most textbooks. The origin of this 
important relation is unitarity~\cite{GOTT,BERRYMOUNT,MILLER}. 
It can be derived assuming {\em only}
unitary evolution by the corresponding Hamiltonian. Alternatively,
one can see it by deriving a first order current conservation differential
equation that is satisfied by $\Delta$~\cite{SHULMAN}. 
For given boundary conditions,
in the setting of quantum mechanics, equation~\pref{unitarity} is essentially
established by making use of the existence and uniqueness theorem
for the corresponding differential equation\footnote{
A particularly elegant approach is also described in~\cite{GOTT} (and
see references therein), where
the authors use the uncertainty relation, and a simple
classical phase space probability estimate to derive~\pref{unitarity}.
}.

While equation~\pref{unitarity} makes it easy to compute the determinant
of the propagation operator, we need to be careful about one aspect of this
computation. In differentiating the classical action, we need to keep
in mind that one holds $t$ fixed, {\em not} $E$, which in turn implicitly
depends on $t$, as well as $v_1$ and $v_2$. This requires some mild
juggling of partial derivatives, that we show in Appendix D. The result 
is, for the canonical momentum,
\bb\label{canmom2}
\PP\equiv \lk(\frac{\del S_{cl}}{\del v_1}\re)_{t,v_2}=
\frac{\Sigma}{\sqrt{2\alpe}} |E| 
\frac{\sqrt{1-(v_1/v_c)^3}}{v_1^2}\ ,
\ee
as the reader may check with respect to $\Pi_v$
(see~\pref{Piv} and~\pref{vdot2}). And, more importantly,
\bb\label{delta}
\Delta=
\lk(\frac{\del \PP}{\del v_2}\re)_{t,v_1}=
\frac{\Sigma}{\le}
\lk(\frac{|E|\pm 2 G \omega v_1^3}{v_1^2 \sqrt{1-(v_1/v_c)^3}}\re)
\lk(\frac{|E|\pm 2 G \omega v_2^3}{v_2^2 \sqrt{1-(v_2/v_c)^3}}\re)
\frac{|E|}{\sqrt{32} G \omega^2}
\II^{-1}\ ,
\ee
with
\bb\label{ii}
\II\equiv \int_{v_1}^{v_2}\ dv\ 
\frac{v \lk(1+G v^3\re)}{\lk(1-(v/v_c)^3\re)^{3/2}}\ .
\ee
This last integral can in turn be written down
in terms of hypergeometric functions. We note that it is 
finite as $v_2\rightarrow 0$. This expression, being
the real part of the propagator, tells us 
about a physical amplitude as a function of the 
corresponding boundary data.

Let us look at~\pref{delta} for the two interesting cases:

\begin{itemize}

\item For the bound state scenario, we have 
\bb\label{deltabnd}
\Delta_{bnd}\simeq 2^{1/6} 3 
\frac{\Sigma}{\le} \omega^{1/3} \frac{G^{1/3}}{|E|^{2/3}}
\frac{\lk(\omega\mp |E|\re)^{4/3} 
\lk(\pm E^2-2\omega |E|\re)}{E^2\mp4\omega |E|+4\omega^2}
\frac{1}{v_2^2}\ ;
\ee
which is divergent as $v_2\rightarrow 0$.

\item For scattering, we get
\bb\label{deltascatt}
|\Delta_{scatt}|\sim \frac{\Sigma}{\le}
\frac{E^2}{4\sqrt{2} G \omega} \frac{1}{v_1 v_2^2 |v_c|^3}\ .
\ee

\end{itemize}

\subsubsection{The DeWitt term}

It was shown by DeWitt that, in evaluating the path integral
for a particle in curved space, coupled to a gauge field, and 
an arbitrary potential, the resulting propagator does not satisfy
the Schr\"{o}dinger equation for the corresponding Hamiltonian. 
To assure that $\GG(v_1,v_2,t)$ is the correct
propagator for the Hamiltonian~\pref{hamil}, DeWitt showed that we
need to add a term $S_{dW}$ as in~\pref{propres}, given by\footnote{
In this approach, the constraint can be imposed
on the Hilbert space, instead of being solved for from the outset.
See the discussion about this in Appendix E.}
\bb
S_{dW}=\int d\tau\ \frac{\RR}{12}
=\frac{1}{16} \frac{\sqrt{2\alpe}}{\Sigma} \frac{1}{|E|}
\int_{v_1}^{v_2}\ dv\ 
\frac{1+2 G v^3\lk(3+G v^3\re)}
{\lk(1+G v^3\re)^2 \sqrt{1-(v/v_c)^3}}\ ,
\ee
where $\RR$ is the scalar curvature associated with the metric
\bb
g_{vv}=\Sigma \frac{\Omega^2}{\pi \alp v^2}\ ,\ \ 
g_{tt}=-\Sigma \frac{v^2 \Omega^2}{8\pi^3 \alpe \alp}\ ,
\ee
that appears in our Hamiltonian. The physical meaning of this curvature
term has long been a mystery\footnote{
In his classic book on path integration, Shulman 
writes ``If you like excitement, conflict, and controversy, especially
when nothing very serious is at stake, then you will love the history
of quantization on curved spaces.'' Our problem can certainly
be attributed by some of these qualifications.
}. The puzzle arises when one
notices that this term carries more powers in $\hbar$ (which is
not easy to see when one sets $\hbar=1$).
Restoring powers of $\hbar$ in the exponent of~\pref{propres}, we should 
have written
\bb
\frac{1}{\hbar}\lk(S_{cl}+\frac{\hbar^2}{12}\int\ d\tau\ \RR\re)\ .
\ee
In our problem, we will instead
count powers of $\hbar$ through powers of the
coupling $G$. Now, let us look at the two cases of interest:

\begin{itemize}
\item For the bound state problem, we get
\bb\label{sdW}
S_{dW}^{bnd}=-\frac{\sqrt{\pi}}{8\sqrt{2}} \frac{\Gamma(4/3)}{\Gamma(5/6)}
\frac{\le}{\Sigma}
\frac{1}{|E|} G^2 v_c^7\ \SS\ ;
\ee
where we define
\bb
\SS\equiv 
\lk(\frac{1}{x^2}\re) \mbox{}_2F_1\lk(\frac{1}{3},2,\frac{5}{6},-x\re)
+\lk(\frac{12}{5} \frac{1}{x}\re) 
\mbox{}_2F_1\lk(\frac{4}{3},2,\frac{11}{6},-x\re)
+\lk(\frac{32}{55}\re) \mbox{}_2F_1\lk(2,\frac{7}{3},\frac{17}{6},-x\re) \ ,
\ee
with
\bb\label{xdef}
x\equiv G |v_c|^3\ .
\ee

\item For the scattering scenario, we have
\bb
S_{dW}^{scatt}\simeq \frac{1}{128 G} \frac{\le}{\Sigma} \frac{1}{|E|}
\lk(\frac{2 |E|}{\sqrt{\pm\omega |E|-\omega^2}} \frac{1}{\sqrt{v_1}}
+\sqrt{G} v_2\re)\ ,
\ee
which interestingly goes to zero for large $v_1$ and small $v_2$.

\end{itemize}

\section{The bound state problem}

In this section, we focus on the bound state scenario,
with two possible orientations for the closed string winding.
We set $v_1=v_c$ and choose $v_2$ small throughout.

To determine the spectrum within the Bohr-Sommerfeld approximation, we use
the condition
\bb\label{born}
S_{tot}^{bnd}=\pi N\mbox{  with  }N\gg 1\ ,
\ee
with
\bb\label{stot}
S_{tot}^{bnd}=S_{cl}^{bnd}+S_{dW}^{bnd}\ ,
\ee
evaluated for the bounded motion $v\in\{0,v_c\}$. $N$ is taken as a large
integer. The action in~\pref{stot} is, remarkably, finite
and corresponds to a non-zero phase picked up
by the closed string during a round trip.
The left hand side of~\pref{born} is a function of $E$,
$\omega$, $G$, and $\Sigma/\le$, and this equation then gives
us the quantization of $E$, and hence of $Q_E$.
There are several issues with regards to this statement that are unusual.
First, the infinite redshift at the horizon $v_2\sim 0$ implies that,
while the string takes finite proper time for a round trip between
$v_c$ and $v_2\sim 0$, it would take infinite time in the time
variable $t$. The dynamics is however novel in that
that the closed string gets {\em repulsed} from
the horizon. Furthermore,
the limit $v_2\rightarrow 0$ is problematic for finite $M$ in view
of~\pref{curvec}.
These issues are addressed in
detail in Appendix E and the Discussion section. 
The reader concerned about them
is urged to consult the appendix at this point.

While we have an analytical expressions for the level spectrum given 
by~\pref{born}, along with~\pref{sclbnd} and~\pref{sdW},
it is instructive to look at
physically interesting limits, so as to write more
transparent expressions. Two cases stand out
\bb\label{subass}
\lk\{
\begin{array}{lll}
|E|\ll \omega & \Rightarrow Q_E\ll \omega \frac{\Sigma}{4\pi \alpe} 
& \Rightarrow x\sim 
\frac{|E|^2}{4\omega^2}\ll 1 \\
|E|\gg \omega & \Rightarrow Q_E\gg \omega \frac{\Sigma}{4\pi \alpe} 
& \Rightarrow x\sim 
\frac{|E|}{4\omega}\gg 1 \\
\end{array}
\re.\ ,
\ee
where $x$ was defined in~\pref{xdef}. Physically, these limits
correspond respectively to energies much below and above the threshold
of creating the corresponding wound closed string in Wound String theory.

We then find
\bb\label{bnd1}
S_{cl}^{bnd}=
-\frac{\Sigma}{\le} 2^{7/6} \pi^{1/2}
\frac{\Gamma(2/3)}{\Gamma(1/6)}
\lk\{
\begin{array}{ll}
G^{1/3} \omega^{2/3} |E|^{1/3} & x\ll 1 \\
\frac{1}{2} G^{1/3} \omega^{1/3} |E|^{2/3} & x\gg 1 \\
\end{array}
\re.\ ;
\ee
\bb\label{bnd2}
S_{dW}^{bnd}=
-\frac{\le}{\Sigma} \frac{\pi^{1/2}}{2^{19/6}}
\frac{\Gamma(4/3)}{\Gamma(5/6)}
\lk\{
\begin{array}{ll}
\frac{1}{2} \frac{1}{G^{1/3} \omega^{2/3} |E|^{1/3}} & x\ll 1 \\
\frac{1}{G^{1/3} \omega^{1/3} |E|^{2/3}} & x\gg 1 \\
\end{array}
\re.\ .
\ee
The expansion parameter in each limit is then the same for
both expressions. For $x\ll 1$, the expansion in $\hbar$ becomes an
expansion in $G |E| \omega^2$; while for $x\gg 1$, it becomes
$G |E|^2 \omega$. Putting~\pref{bnd1} and~\pref{bnd2} together in~\pref{stot},
we find
\bbb
Q_E\le & \sim \lk(\frac{\le}{\omega \Sigma}\re)^2 \frac{N^3}{G} 
& \mbox{  for  }
x \ll 1\mbox{  with the two winding possibilities degenerate} \label{finspec1}\\
Q_E\le & \sim \sqrt{\frac{\le}{\omega \Sigma}} 
\frac{N^{3/2}}{\sqrt{G}} 
& \mbox{  for  }
x \gg 1\mbox{  for negative winding only}\label{finspec2}
\eee
The DeWitt term corrects only the
numerical coefficients in front of these expressions;
the scaling with respect to all physical parameters of
the theory is fixed by the leading term $S_{cl}^{bnd}$.
From~\pref{subass}, \pref{finspec1} can be trusted
for $N\ll N_{max}$ with $N_{max}\equiv \omega (\Sigma/\le) G^{1/3}$; 
while~\pref{finspec2} can be trusted
for $N\gg N_{max}$. 
Beyond these two asymptotic 
regimes, one needs to use the full analytic forms given by~\pref{sclbnd},
\pref{sdW} and~\pref{born}, which we plot in Figure~\ref{specfig}. Note
also that the spectrum does not have any explicit dependence on $M$.

From Figure~\ref{specfig}, we see how the density of levels increases
as we approach the threshold energy from below, in the case
of a positively wound closed string. Much below the threshold,
where the two windings correspond to degenerate levels, 
the energy scales as $N^{3}$. For
the negatively wound case, we go past the threshold energy with a peculiar
$N^{3/2}$ scaling with the level number. It also appears that the
negatively wound states have always higher energies than the positively
wound ones for same $N$.
If we could trust the spectrum for low enough energies, we would also be
identifying a ground state with
energy slightly above zero, scaling with respect
to the parameters of the theory as shown in
equation~\pref{finspec1}. Finally, we note that
the spacing of energy levels increases with $N$
as $\delta Q_E/\delta N\sim N^2$ for positive winding and for energies 
well below the threshold, and as
$\delta Q_E/\delta N\sim \sqrt{N}$ for negative winding and high energies. 

Even with the full form of the quantization condition depicted
in the Figure, we need to be careful in trusting the spectrum, given
the restrictions on the spacetime imposed by~\pref{curvcond}, \pref{curveb}
and~\pref{curvec}. In particular, we need to make sure that the region of space
explored by the closed string lies within the area delineated
by the curves of Figure~\ref{validity}. For example,
for~\pref{finspec1}, we need 
$\alpe/(4\pi\Sigma^3)\ll Q_E\ll \omega G^2 \Sigma/(4\pi \alpe)$ for
$G\ll 1$ (corresponding to the lower small triangular 
region in Figure~\ref{validity});
$\omega/(4\pi\Sigma)\ll Q_E\ll \omega \Sigma/(4\pi \alpe)$ for
$1\ll G \ll \sqrt{M} \le/\Sigma$; and 
$\omega \Sigma G^2/(4\pi\alpe M)\ll Q_E\ll \omega \Sigma/(4\pi \alpe)$ for
$\sqrt{M} \le/\Sigma \ll G$; while for the second expression, one needs 
$\omega \Sigma/(4\pi\alpe)\ll Q_E\ll \omega \Sigma M^2/(4\pi \alpe G^4)$.
One should not make too much of these additional conditions; the only relevant
point is that there are certain restrictions and we can arrange a large
hierarchy of energies where our expression for energy levels can be trusted
by changing $G$, $M$, and $\Sigma/\le$. This is shown
in the figure above 
with a shaded region. The important points
are that (a) the energy levels scale inversely with the
coupling $G$, and hence are non-perturbative in character; (b)
the scaling with the integer $N$ is large enough that the higher levels 
will be spaced more and more; 
(c) that the spectrum is regular in the limit $M\rightarrow\infty$;
a regime which circumvents complications that affect the dynamics near
$v\sim 0$; (d) that
there exists states with the `wrong' winding orientation that
cannot escape to infinity and yet explore energies above the threshold;
and (e) that, for positive winding,
the transition point between freed closed strings and
NCOS states is the threshold energy. As pointed out earlier, this
threshold energy appears with the correct additional $1/2$ factor
that has already been attributed to the closed string tension~\cite{KLEBMALDA}.

Most interestingly,
the $M\rightarrow \infty$ limit,
with all other parameters of the theory held fixed, is regular;
this renders the whole of space, from $v=0$ to $v\rightarrow\infty$,
reliable for our computations. In this strict limit, the spectrum is in 
principle good for all energies as to the issues pointed out in
Section~\ref{valsec}, but
except for issues having to do with reliability
of a WKB-like approximation. These latter matters are controllable
and are discussed in Appendix E as well as the Discussion section.
For finite but large $M$, we argue in the Discussion section that this
spectrum may constitute a first order estimate of energy levels
in an expansion in $1/M$.

Let us next look at the propagators in the bound state problem.
We emphasize that the
region explored by the bound state motion is not confined to the throat
$G v^3<1$, but can involve the whole of the
space; as seen from equation~\pref{vc},
for both windings, $v_c$ ranges from $0$ to $\infty$, for the allowed
values for $E$ as given in~\pref{table}.
Expression~\pref{propres} for the
propagator tells us about the probability of propagation in the $v$
direction (see footnote at the beginning of Section~\ref{quantumeff} 
with regards to the measure); 
in the dual NCOS theory, this presumably translates to probability 
for a corresponding NCOS state to live and `breath' for some time;
location in the $v$ direction being mapped onto size of a soliton
in the NCOS theory. Note also that these states carry zero total
momentum in the $y$ direction. 
In order to interpret~\pref{propres} as a probability amplitude,
we need however to eliminate all instances
of $E$ in the propagator in favor of $\Delta t$.
This is identical to the
situation that arises say in $AdS_3$ where one tries to 
find the correlators of two operators in the conformal field theory
by looking at a geodesic motion in the bulk, \ie\ at a propagator
in the bulk. We would then need to invert~\pref{tint}
to write $E$ as a function of $\Delta t$; a difficult task that we will
not be able to do. The problem here is that the motion explores the
space all the way near the center; whereas in the $AdS_3$ case for example,
the corresponding geodesic motion lives near the boundary, and leads to
a trivial relation $t\sim 1/E$. We cannot do this here as the motion is
not confined near a boundary, about which we would expand. For the
case of the scattering solution that we discuss in the next section,
the problem will be the same, as the closed string falls from infinity
all the way to the horizon.
In an effort to unravel the dynamics, we will then look at the two limits
$|E|\gg\omega$ and $|E|\ll\omega$, and expand the relations between 
$t$ and $E$ in these regimes only.

The expression given by~\pref{deltatbnd} 
is not enough and one needs the subleading
term to extract the energy dependence. This is a subtle limit
to take, which we do carefully by taking the limits on energy
first, then integrating~\pref{tint}. The procedure is outlined briefly
in Appendix C. The result are simple relations (for $v_1=v_c$)
\bb\label{eoft1}
|E|^{1/3}= 2 \sqrt{G}\omega \sqrt{\frac{D_1 v_2}{1-T v_2}}\mbox{     for  }
|E|\ll\omega\ ;
\ee
and
\bb\label{eoft2}
|E|^{1/3}= 4 G\omega \frac{D_2 v_2}{1-T v_2}\mbox{     for  }
|E|\gg\omega\ .
\ee
$D_1$ and $D_2$ are numerical constants given by
$D_1=\sqrt{\pi} \Gamma(2/3)/\Gamma(1/6)$ and $D_2=2 D_1$.
And $T$ is defined as
\bb
T\equiv \frac{\Delta t_{bnd}}{\sqrt{8\pi^2 \alpe}}\ ,
\ee
\ie\ the time in NCOS string units. In these expressions,
we have also expanded in powers of
$v_2/v_c\ll 1$. 
These are not the typical relation $|E|\sim 1/T$; as $v_2\rightarrow 0$,
and $T\rightarrow\infty$, with the product $T v_2$ getting close to $1$,
we have $E$ remaining finite. Note also that $v_2$ is the IR cutoff in
the dual NCOS theory, so that the $T v_2$ is a natural combination.
In particular, we have the requirement $T v_2\leq 1$; \ie\ we should not
probe dynamics for times longer than the one set by the IR cutoff.
We have also verified these asymptotic expansions numerically, and found they
are excellent approximations to the exact form within the regimes
of interest\footnote{
Note that the exponent in~\pref{propres} could in principle be used
to extract the spectrum of states in an alternative way. 
One would Fourier transform in the time variable $t$ dual
to energy in the NCOS theory; then, look for poles in energy. This is 
technically a much
more involved approach in this case for obvious reasons.}.

The interesting quantity
to look at is the real part of the propagator~\pref{propres}, 
$|\GG|^2\sim |\Delta|$. 
Let us focus on the two interesting regimes in expression~\pref{deltabnd}:
\bb\label{deltabnd2}
|\Delta_{bnd}|\sim \lk\{
\begin{array}{ll}
\omega^{2/3} \frac{\Sigma}{\le} G^{1/3} \frac{|E|^{1/3}}{v_2^2} & \mbox{  for  }
|E|\ll \omega \\
\omega^{1/3} \frac{\Sigma}{\le} G^{1/3} \frac{|E|^{2/3}}{v_2^2} & \mbox{  for  }
|E|\gg \omega
\end{array}\re.\ .
\ee
Eliminating $E$ between~\pref{eoft1}, \pref{eoft2} and~\pref{deltabnd2}, we get 
\bbb
|\GG|^2 & \sim
\omega \frac{\Sigma}{\le} \frac{\sqrt{G}}{v_2^{3/2}}
\frac{1}{\sqrt{1-T v_2}} & \mbox{  for  }
v_1^3\ll \frac{1}{G} \label{deltabndf1} \\
|\GG|^2 & \sim
\omega \frac{\Sigma}{\le} G \frac{1}{1-T v_2} & \mbox{  for  }
v_1^3\gg \frac{1}{G} \label{deltabndf2}
\eee
Note that, in these expressions, $v_c$ is the initial position
$v_1=v_c$; and we need to be careful to take $v_2\ll v_c$, in particular
for~\pref{deltabndf1}. Also, the dependence on $T$ includes
contribution from $v_c$; so that our expression is a function of $v_2$
and $T$ independently, with fixed initial position of propagation at
the classical turning point. We interpret these expressions as
the probability for a corresponding configuration in NCOS theory to 
spread in size from its most compact form to a size $1/v_2$ in a time $T$.

\section{The scattering problem}

The scattering solution exists only for positive winding. The
closed string starts at $v$ large, where perhaps it can be used to
define an asymptotic on-shell state within Wound string theory, 
and falls to $v\rightarrow 0$, while decelerating as it approaches
the horizon. It would then bounce back to infinity. For fixed initial
position $v_1$, this process takes finite proper time, as seen
from~\pref{deltatauscatt}. In the time variable $t$, which we associate with
the NCOS theory, this process seems to take infinite time. This presents
a problem in trying to interpret the process as scattering within the
dual NCOS dynamics.

The natural suggestion of this setup is that this solution
describes a closed string in Wound String theory scattering off the
bound state of D-strings and fundamental strings. This correlates
well with the fact that strings of only positive winding can undergo
this process. Given that $v_c$ for the
positively wound bound state approaches infinity as 
$Q_E\rightarrow \omega \frac{\Sigma}{4\pi\alpe}$, which is the threshold
of creating a wound closed string (see equation~\pref{vc}), we may expect that
in the dual NCOS theory the scattering process is encoded 
holographically by the insertion of the appropriate local
vertex operators at large values of $v$, in the UV. Hence in this case,
$\Delta$ given by~\pref{deltascatt} would tell us about the amplitude
for the incoming closed string to break up on the D-strings, breath
for a while through a resonance (with zero net momentum in the $y$ direction), 
and leave the NCOS theory. We will carry on in computing this amplitude,
keeping in mind that we do not have a satisfactory resolution of the
issue that this process yet appears to take infinite times in the variable $t$ .
We consider in this section $v_1$ large and $v_2$ small throughout.

Using equation~\pref{deltatscatt}, we solve for $E$ and find
\bb\label{eoftscatt}
|E|=4 G\omega \frac{v_1 v_2^2}{\lk(T v_2-1\re)^2}\ .
\ee
And taking the limit $|E|\gg\omega$ in~\pref{deltascatt}, we get 
\bb
|\Delta_{scatt}|\sim \frac{\Sigma}{\sqrt{2\alpe}} |E| \frac{1}{v_1 v_2^2}
\sim \omega \frac{\Sigma}{\le} \frac{G}{\lk(T v_2-1\re)^2}\ ,
\ee
where in the last step, we used~\pref{eoftscatt} to eliminate $E$. 
In particular, the divergences from $v_1$ and $v_2$, the IR and UV cutoff
cancel. Another interesting aspect of this result is the linearity
in the NCOS coupling. We can interpret this propagator as an amplitude
in NCOS theory associated with a local insertion of a single vertex operator
corresponding to the closed string in the parent Wound string theory;
with $v_2$ being interpreted as the size of the final state,
presumably a non-perturbative coherent state,
in the NCOS theory. Yet, at strongly coupling, this amplitude scales
linearly with the NCOS coupling. For the same technical
reasons encountered above in inverting $t$ as a function of $E$,
the bound state spectrum is difficult to unravel from the scattering
propagator.

For the reader's convenience, we write in this regime
the leading phase of the propagator. We find, after eliminating $E$
\bb
S_{cl}^{scatt}\sim 2 \omega \frac{\Sigma}{\le} G \frac{v_1 v_2}{T v_2-1}\ ,
\ee
Hence, there is a suggestive correlation between the UV and IR cutoffs, as
they appear in the product $v_1 v_2$ that perhaps could be held finite
in a controlled limit.

\section{The holographic duality for NCOS}
\label{holosec}

In this section, we collect several observations suggesting that we may
associate a screen, located at finite $v$
in the bulk space, with the dual NCOS theory, that
holographically encodes dynamics in the whole of space. Consider
an observer siting at the non-commutativity throat
\bb
v^3_o=\frac{\kappa}{G}\ ,
\ee
where $\kappa$ is an arbitrary numerical constant.
If this observer was to measure, locally, energy
of our projectile, it would be
\bb\label{locale}
 \varepsilon=\lk.|G_{00}|^{1/2} \frac{dt}{d\tau}\re|_{v_0}
\sim |E|+\mbox{constant}
\ee
\ie\ all instances of $G$ disappear from~\pref{locale}. 
Furthermore, evaluating the metric~\pref{metric} at $v_0$, we get
\bb
\lk.G_{\mu\nu}\re|_{v_o}=\frac{G_s^2}{\gs} \sqrt{\kappa}\sqrt{1+\kappa}\ .
\ee
While the dilaton is
\bb
\lk.e^\phi\re|_{v_o}=
G_s \frac{1+\kappa}{\sqrt{\kappa}}\ ;
\ee
(see also~\pref{Gs}).
And from the worldsheet action of the closed string in this region, the
effective string length at this point is
\bb
\alp|_{v_0}=\frac{\alp}{|G_{00}|_{v_0}}=
\frac{\alpe}{\sqrt{\kappa}\sqrt{1+\kappa}}\ .
\ee
These four expressions are, up to a numerical constant logged by
$\kappa$, the NCOS energy scale, metric, (closed string) coupling, 
and string scale respectively, 
in {\em flat space}\footnote{In particular, we remind the reader of the
NCOS map
$G_{\mu\nu}=G_s^2/\gs$~\cite{SWNC,GMSS,KLEBMALDA,GMMS}. We may also refer
to this as the Wound String theory map adopted to the NCOS setting.}.
Hence, it is natural to think that the dual NCOS
theory to our spacetime `sits' at an energy scale
$v\sim v_0$, which is
the throat region of the spacetime geometry depicted in Figure~\ref{throat1}. 
It is the region
analogous to the `boundary'
of space in the Maldacena scaling limit in the cases with zero
B-field.
Note however that, for the bound state solutions, the closed string
reaches up to an extent $v_c$, which can get larger than $v_0$ for high
enough energies. And it then appears that the whole of space
is available for holographic encoding.

Finally, it is important that, at $v\rightarrow \infty$, within the
same framework, we also recover
the conventionally different scaling limit of Wound String theory
$g_{\mu\nu}=\lk(-1,1,\delta,\delta,\ldots\re)$,
$\alp\sim \delta$, and
$\gs\sim \frac{1}{\sqrt{\delta}}$,
as shown explicitly in~\cite{DGK2}, with $\delta\sim 1/v^3$.

\section{Discussion}

Two aspects of our setup need more elaboration. The
first has to do with the fact that our formalism neglects quantum fluctuations
beyond the ansatz we used. We have been discussing the dynamics
of the center of mass of the closed string, with zero angular
momentum on the seven sphere. We showed that this ansatz can be 
classically solved for self-consistently.
We studied the
quantum mechanics with sum over paths restricted to this ansatz;
yet the string has available to itself a much larger phase space for
quantum fluctuations than that of a point particle in one dimension. 

If we were to imagine how the spectrum we calculated would look like
had we solved the full problem exactly, we may expect a tower
of levels associated with the bounded motion of the center of mass,
and, superimposed on this, another level spectrum associated with fluctuations
on the string about the center mass. We may expect correlations
between the spacings of these two classes of energy levels.
The spectrum we have computed 
must then correspond to approximating part of this
full spectrum. We should expect two additional effects: first, additional
levels towering on each energy level we predict, corresponding to
vibrations of the string itself; and second, possibly additions and
corrections to the levels we computed from fluctuations of the zero modes that
take us away from the ansatz; corrections, in particular, that come from the
zero point energy of other oscillators.
We suggest confining the results
to large values of $N$, where the energy level spacings become large
parametrically with $N$. Other fluctuations being independent of this,
they become of less importance for the higher levels.
Also, our WKB-like approximation is naturally improved for large $N$.

The other issue of concern has to do with the
part of the motion of the closed string near the center $v\sim 0$.
The problem here is that, even as we push
$\Sigma$ and, more importantly,
$M$ large enough so as to trust the computation to small
values of the radial coordinate, we will eventually venture into a regime
where the calculation breaks down near the repulsive horizon at $v=0$
if $M$ is finite;
and the question is with regards to the effect of this forbidden region on the
overall picture of the dynamics we painted.

As mentioned earlier, the strict limit $M\rightarrow\infty$, 
with $G$, $\Sigma$, and $\le$ fixed, is a well-defined regime with all
physical measurables remaining finite. This is strong evidence that
we have identified the coupling of the theory $G$ correctly as in~\pref{cpl}.
More interestingly, in this regime, the issues that plague dynamics in the
$v\sim 0$ region do not exist, as the geometry is reliable from $v=0$
to $v\rightarrow\infty$. The spectrum computation
is then controlled, up to the standard issues associated
with the WKB-like approach discussed in Appendix E, 
and the restriction of the phase
space to the $v-t$ plane discussed above; both matters are effects we can
understand and control by taking the level number large. Note also that, 
for $M\rightarrow\infty$, there is a possibility that,
the WKB-like calculation being coincidentally exact as it sometimes can be,
and with the help of supersymmetry in controlling zero point energies of
other oscillators in the system, we may have
part of the exact spectrum of a string theory at strong coupling.

For finite $M$, the picture is more problematic.
First, let us note that, in principle we can extend the calculation 
to smaller values in $v$ by applying duality transformations. But this
is bound to postpone the question instead of answering it;
either into an S-dual picture, or strongly coupled IIA theory
(after applying a T-duality). We instead will attempt to speculate within 
IIB theory as to the effect of the center of the geometry
on the dynamics in the bulk for finite $M$.

Near the origin of the radial coordinate,
the local string coupling is becoming big
with smaller values of $v$,
as can be seen from Figure~\ref{validity}.
Furthermore, as the string falls onto the center,
it is decelerating and its multiply wound strands are getting 
squeezed into a smaller area of space. As it spends longer and longer
proper times 
in this region, we expect string interactions to become more and more
important for this part of the motion. In the first quantized formalism
we adopted, we should then consider processes where the closed string
splits and joins, while conserving total winding number. One could describe
these processes by introducing this surgery by hand in the center of
the geometry, and representing the entire process by gluing the corresponding
trajectories near the center, as classical solutions that  
interpolate between the various solutions we have written.
The effect of these interactions is subleading to the overall
dynamics, weighed by powers of the local string coupling, and correspondingly,
as can be seen from equation~\pref{dil}, by powers of $1/M$, with $M$ being
finite but large. This
suggests that we may
describe the picture by the spectrum we have computed,
by allowing possibly copious transitions between the various energy levels
across different winding sectors. In the dual NCOS picture, this would
correspond to non-perturbative states that can decay amongst each other,
with their masses being estimated well to first order in $1/M$ 
by the quantum mechanical spectrum
computed. This is a first look into
what appears to be a complex and rich non-perturbative
dynamics in NCOS theory at finite $M$. 
Somewhat similar phenomena may be at work in
the SL(2,R) WZW model (see discussion in~\cite{MALDAOOG1}).

Hence, the picture we have is that there exists states in the NCOS at
strong coupling,
with masses that scale inversely with the coupling. These objects,
which carry zero momentum in the $y$ direction, may have time dependent
sizes; perhaps similar to breathing modes. There can be created in scattering
processes involving a closed string of Wound String theory, perhaps
represented as a vertex operator insertion in the NCOS theory.
In certain asymptotic regimes, we wrote propagators
of the NCOS theory, as a function of time and
size. A general observation was that these have poles at
times of order the final size $T\sim 1/v_2$. 
A linear scaling in the coupling, along with certain non-trivial cancelations
of potential divergences in the UV, are some of the interesting
features we encountered.
To address these issues, it seems important that we understand the 
infinite redshift effect at the horizon tied to
our choice of what we call energy or time.

Another interesting matter has to do with our observation that,
in the $M\rightarrow\infty$ limit, we have a well-defined background
geometry for all $v$. This may be viewed as a mechanism whereby
non-commutativity, through the introduction of the throat, 
regulates the singularity that would otherwise arise at $v=0$.
The potential trouble gets replaced in the center with a patch of
spacetime which is almost flat, much like a similar mechanism proposed
in~\cite{JPJ}. In this case, the regulator is non-commutativity in the infinite
$M$ limit. It would be interesting to pursue this line of thought further
in other systems.
 
Beyond resolving some of these open questions, 
let us also comment briefly on potentially interesting
future directions.
First, understanding the dynamics of putting momentum along the $y$
direction, as well as vibrations on the closed
string, is important. The problem is not intractable, as one can adopt
an approach of a perturbative expansion about the center of mass
dynamics (if needed) to gain at least a hint of how the spectrum evolves. 
It is useful however, at this stage, to understand and include the
effects of the RR fields. This may necessitate a more controlled approach
by using the symmetries in the theory, in particular supersymmetry.
There are interesting qualitative similarities between our problem
and the SL(2,R) WZW model analyzed in~\cite{MALDAOOG1}. In understanding
that system, a crucial role is played by spectral flow, in a system
that is an exactly solvable CFT. It would be interesting to see
whether one can find an analogous operation
that may be used to study the NCOS problem in a more
mature manner. In this respect, an important simplification may be
achieved by working in the strict limit $M\rightarrow\infty$.

Another line of thought is to attempt to understand dynamics of D-branes
in this background. In particular, D-string motion may entail interesting
information, the setting being S-dual to the one considered. And it
would be desirable to better understand how one should think of
holography in this context. It appears this system may be used to
test and study the Holographic duality beyond
the near horizon scaling limit; within Wound String
theory, the energy regime we consider is a superset of the conventional
regime associated with Holography~\cite{DGK2}. In this respect, our analysis
demonstrate that previous attempts to identify a criterion for
Holography are too restrictive~\cite{BOUSSOCONJ,AGC,CFUNCTION}, 
as they confine themselves to
short wavelength, point-like probes; geodesics, which are only part of the
story in a picture involving strings. We hope to report on this in
an upcoming work.

\section{Acknowledgments}
I thank P. Argyres and T. Becher for discussions. I am grateful
to K. Gottfried for making part of his manuscript of his upcoming
textbook on Quantum Mechanics available for reading. 
This work was supported in part by a grant
from the NSF.

\section{Appendix A: The $(N,M)$ string solution}

In this appendix, we present, for the convenience of the reader,
the background fields generated 
by the $(N,M)$ string, and the associated NCOS limit. 
The parameter space in the parent IIB theory is given by
\bb
\lk\{N,M,\gs,\alp,\Sigma\re\}\ .
\ee
The metric can be found in~\cite{SL2Z,VVSNCOS}
\bb
ds_{str}^2=\gs \sqrt{\frac{K}{L}} \lk \{
A^{-1/2} \lk ( -dt^2+\Sigma^2 dy^2 \re ) 
+A^{1/2} \lk ( dr^2+r^2 d\Omega_7^2 \re)\re \}\ .
\ee
With the NSNS fields
\bb
B_{ty}=\gs^2 \Sigma \frac{N}{M} A^{-1} L^{-1/2}\ ;
\ee
\bb
e^\phi=\gs A^{1/2} \frac{K}{L}\ .
\ee
And the RR fields
\bb
A_{ty}=\Sigma \lk(A^{-1}-1\re) L^{-1/2}\ ;
\ee
\bb
\chi=\frac{N}{M} A^{-1} \frac{A-1}{K}\ .
\ee
The various variables appearing in these expressions are defined as
\bb
A\equiv 1+\frac{q^6}{r^6}\ ,\ \ 
K\equiv 1+A^{-1} \lk(\frac{N \gs}{M}\re)^2\ ,\ \ 
L\equiv 1+\lk(\frac{N\gs}{M}\re)^2\ ,
\ee
with
\bb
q^6\equiv \frac{32 \pi^2}{\gs^2} M \alp^3 L^{1/2}\ .
\ee
This solution has manifest SL(2,Z) symmetry.

The NCOS limit is obtained by
\bb
\alp\rightarrow 0\ ,\ \ 
\gs\alp=\alpe G_s^2\mbox{  fixed  }\ ,\ \ 
\mbox{and  }U\equiv \frac{r}{\alp}\mbox{   fixed}\ ,
\ee
with $\alpe$ defined by this expression, and $G_s$ defined below.
In the main text, we have also performed a coordinate change
\bb
U^2\equiv \frac{V}{\le^3}\equiv \frac{M^2}{(32\pi^2)^3 G^3 \alpe} v\ ,
\ee
to put the metric in a form conformal to $AdS_3\times S^7$.
Furthermore, $v$ becomes energy scale in the UV-IR relation
$v\sim \mu_{NCOS}$.

In this limit, the dual theory is 1+1 dimensional NCOS theory,
with (closed) string coupling
\bb\label{Gs}
G_s=\frac{M}{N}\ ,
\ee
which is more conveniently written as
\bb\label{cpl}
G\equiv \frac{G_o \sqrt{M}}{32\pi^2}\ ,\ \ G_o\equiv \sqrt{G_s}\ .
\ee

\section{Appendix B: The path integral}

We need to setup a path integral for the propagator 
of our quantum mechanical system, and evaluate it,
subject to the first class constraint
\bb
C_1=\frac{2\pi^2}{\Sigma^2} \frac{\alp^2}{\Omega^2} v^2 \Pi_v^2
-\frac{16 \pi^4 \alpe }{\Sigma^2 v^2} \frac{\alp^2}{\Omega^2}
\lk(\Pi_t\mp\frac{\omega G \Sigma}{2\pi \alpe} v^3\re)^2
+\omega^2 \frac{v^2\Omega^2}{16\pi^2\alpe}=0\ .
\ee
The term `first class' refers to the fact that $C_1$ commutes with
the Hamiltonian (in this case, it {\em is} our Hamiltonian). 
The proper approach is to supplement the constraint with
a secondary constraint that is consistent with the equations of
motion~\cite{KOS}. In our case, we choose
\bb
C_2=t-T(\tau)=0\ ,
\ee
where
\bb
\frac{dT}{d\tau}=\lk.\frac{dt}{d\tau}\re|_{v\rightarrow v_{cl}(\tau)}\ .
\ee
The Poisson bracket of the two constraints is then
\bb
\lk\{C_1,C_2\re\}=-2 \lk(\Pi_t\mp\omega \frac{\Sigma}{2\pi\alpe} G v^3\re)\ .
\ee

We now can write down the appropriate path
integral, being careful to start in the Hamiltonian formalism\footnote{
If arranged in the Lagrangian picture in the naive way, 
a path integral in curved space
gives the incorrect measure. This is because of the coordinate
dependent factors multiplying the kinetic term (see for example~\cite{RAMOND}).}
\bb
\GG\sim \int
\DD t\ \DD \Pi_t\ 
\DD v\ \DD \Pi_v\ 
\delta(C_1) \delta(C_2) \sqrt{\lk|\mbox{Det}\lk\{C_1,C_2\re\}\re|}
e^{i \int_0^{\tau_0}d\tau\ \lk(\dot{v} \Pi_v
+\dot{t} \Pi_t-H(t,\Pi_t,v,\Pi_v)\re)}\ .
\ee
Using the well-known identity
\bb
\delta(C_1)\rightarrow 
\frac{1}{2 \lk|\Pi_t\mp\omega\frac{\Sigma}{2\pi\alpe} G v^3\re|} 
\delta\lk(\Pi_t-\Phi_t\re)\ ,
\ee
with $\Phi_t$ given by solving for $\Pi_t$ in $C_1=0$,
the determinant in the measure cancels. We then have
\bb
\GG\sim \int\ \prod_\tau\ dv(\tau)\ d\Pi_v(\tau)\ 
e^{i\sum_\tau\ \delta\tau \lk(\Pi_v \dot{v} +\Phi_t \dot{T}\re)}\ .
\ee
We now perform a change of coordinates from $\tau$ to $t$, using $T(\tau)$
\bb
\int\ d\tau \lk(\Pi_v\dot{v}+\Phi_t\dot{T}\re)=
\int\ dt\lk(\Pi_v \frac{dv}{dt}+\Phi_t\re)\ .
\ee
We then have the expressions given in~\pref{prop} and~\pref{Phit}.

The phase space has been reduced from four to two dimensions. On
this subspace, the evolution operator is $-\Phi_t$, evolving
life in the time variable $t$. The extremum of the integrand in~\pref{prop} 
is at
\bb
\frac{dv}{dt}=-\frac{\delta\Phi_t}{\delta\Pi_v}\ ,\ \ 
\frac{d\Pi_v}{dt}=\frac{\delta\Phi_t}{\delta v}\ .
\ee
The reader may check that these correspond to
the equations of motion~\pref{radialeom} and~\pref{timeeom}, 
subject to~\pref{expconst}.

\section{Appendix C: Some computational details}

In this appendix, we collect a few details the reader
may find useful in checking some of the formula appearing in the main
text.

To compute many of the integrals we encountered, or to write down
their asymptotic behaviors, the coordinate change
\bb
\lk\{
\begin{array}{ll}
\sinh(y)=v^{3/2}/|v_c|^{3/2} & \mbox{  for  } v_c^3<0 \\
\sin(y)=v^{3/2}/|v_c|^{3/2}  & \mbox{  for  } v_c^3>0
\end{array}
\re.
\ee
is useful.
For example, the classical action becomes
\bb
S_{cl}=\lk.\pm \frac{2\sqrt{2}}{3} \frac{\Sigma}{\le}
\omega G \frac{E\pm2\omega}{E}
\sqrt{1-(v/v_c)^3} v_c^2
\mbox{ }_2F_1
\lk(\frac{1}{2},\frac{1}{3},\frac{3}{2},1-\frac{v^3}{v_c^3}\re)\re|_{v_1}^{v_2}\ .
\ee
While the integral given in~\pref{ii} becomes
\bb
\II=-\frac{v_c^3}{6} 
\lk\{
4 \lk(-\frac{1}{v_c^3}-G\re) 
\frac{v^2}{\sqrt{1-(v/v_c)^3}}
+\lk(\frac{1}{v_c^3}+4 G\re) v^2 
\mbox{ }_2F_1\lk(\frac{2}{3},\frac{1}{2},\frac{5}{3},\frac{v^3}{v_c^3}\re)
\re\}_{v_1}^{v_2}\ .
\ee

In finding the asymptotic expansion of~\pref{tint}, we 
rescale $v\rightarrow v/v_c\equiv y$, then note that the numerator of the
integrand becomes $|E|\pm 2\omega G v^3\rightarrow |E|$ for $|E|\ll \omega$;
while it becomes $|E| (1-y^3/2)$ for $|E|\gg\omega$. We then evaluate
the integral in each case in term of hypergeometric functions, which have
well defined expansions for $v_2/v_c\ll 1$.

\section{Appendix D: Calculation of the determinant} 
\label{detapp}

In this appendix, we show the computation of the determinant
$\Delta$ appearing in~\pref{propres}. For this, we need to compute
derivatives of $S_{cl}$ while holding $t$, $v_1$, and $v_2$ 
fixed as needed. This means we have to eliminate $E$
that appears in $S_{cl}$
\bb
\lk\{ S_{cl}(E,v_1,v_2), t(E,v_1,v_2)\re\}
\rightarrow S_{cl}(t,v_1,v_2)\ .
\ee

For example, the canonical momentum to $v$ is given by
\bb\label{canmom}
\PP\equiv 
\lk(\frac{\del S_{cl}}{\del v_1}\re)_{t,v_2}=
\lk(\frac{\del S_{cl}}{\del v_1}\re)_{E,v_2}+
\lk(\frac{\del S_{cl}}{\del E}\re)_{v_1,v_2}
\lk(\frac{\del E}{\del v_1}\re)_{t,v_2}\ ,
\ee
with
\bb
\lk(\frac{\del E}{\del v_1}\re)_{t,v_2}=
-\frac{\lk({\del t}/{\del v_1}\re)_{E,v_2}}
{\lk({\del t}/{\del E}\re)_{v_1,v_2}}\ .
\ee
The expressions on the right hand sides can be derived 
in a straightforward manner using techniques
found typically in thermodynamics textbooks.
The determinant is also given by
\bb
\Delta\equiv
\frac{\del^2 S_{cl}}{\del v_2 \del v_1}=
\lk(\frac{\del \PP}{\del v_2}\re)_{t,v_1}=
\lk(\frac{\del \PP}{\del v_2}\re)_{E,v_1}-
\lk(\frac{\del \PP}{\del E}\re)_{v_1,v_2}
\frac{\lk({\del t}/{\del v_2}\re)_{E,v_1}}
{\lk({\del t}/{\del E}\re)_{v_1,v_2}}\ .
\ee

We now use equations~\pref{sclint} and~\pref{tint} to evaluate these
derivatives. We get
\bb
\lk(\frac{\del t}{\del v_1}\re)_{E,v_2}=
-2\pi \frac{\sqrt{2\alpe}}{E}
\frac{-E\pm 2 G \omega v_1^3}
{v_1^2 \sqrt{1-(v_1/v_c)^3}}\ .
\ee
And
\bb\label{int1}
\lk(\frac{\del t}{\del E}\re)_{v_1,v_2}=
8\pi G \frac{\sqrt{2\alpe}}{E^3} \omega^2
\int_{v_1}^{v_2}\ dv\ 
\frac{v \lk(1+G v^3\re)}{\lk(1-(v/v_c)^3\re)^{3/2}}\ .
\ee

From the classical action, we have
\bb
\lk(\frac{\del S_{cl}}{\del v_1}\re)_{E,v_2}=
\sqrt{2} G \frac{\Sigma}{\le}
\frac{2\pm E}{E}
\frac{v_1}{\sqrt{1-(v_1/v_c)^3}}\ .
\ee
And
\bb\label{int2}
\lk(\frac{\del S_{cl}}{\del E}\re)_{v_1,v_2}=
2\sqrt{2} \frac{G}{E^2} \frac{\Sigma}{\le}
\omega^2 \int_{v_1}^{v_2}\ dv\ 
\frac{v \lk(1+G v^3\re)}{\lk(1-(v/v_c)^3\re)^{3/2}}\ .
\ee

Note that the integrals in~\pref{int1} and~\pref{int2} are identical.
This results in a cancelation in computing~\pref{canmom} that simplifies
the problem considerably. Putting everything together, we get
the expressions given in~\pref{canmom2} and~\pref{delta}.

\section{Appendix E: Comments on the Bohr-Sommerfeld approximation}
\label{bohrapp}

In this appendix, we discuss the merits of the method we used in
deriving the energy spectrum associated with the bounded motion
of the closed string. We also use this opportunity to briefly discuss
the problem from a slightly different angle.

We can think of our quantum mechanical problem in the Schr\"{o}dinger
picture as one corresponding to a Hilbert space spanned
by energy eigenstates
\bb
H |{\Psi}\rangle =\EE |\Psi\rangle\ ,
\ee
with the additional prescription to project onto
the space of states with zero
`energy' $\EE$; \ie\ we apply our constraint directly
on the Hilbert space. We also have
\bb
\Pi_t |{\Psi}\rangle =\frac{\Sigma}{4\pi\alpe} E |\Psi\rangle\ ,
\ee
since $\Pi_t$ commutes with the Hamiltonian $H$.

Our Hamiltonian has the form
\bb
H=f(v)^2 \frac{\Pi_v^2}{2}+\ldots\ .
\ee
The peculiar factor multiplying the momentum is
not a concern, since one can always apply a canonical
transformation {\em without} changing the energy spectrum. For example,
using the generator
\bb
\HH(v,P_v): \lk(v,\Pi_v\re)\rightarrow \lk(W,P_v\re)\ ,
\ee
with
\bb
\HH(v,P_v)=P_v \int_{v_0}^v\ \frac{dv}{f(v)}\ ,
\ee
puts the problem in the standard form
\bb
H\rightarrow \frac{P_v^2}{2}+\VV(W)\ .
\ee

The corresponding Bohr-Sommerfeld quantization rule may be written as\footnote{
Note that we are not quantizing $\EE$, which is fixed to zero by
the constraint; this is a statement of quantization for $E$.}
\bb\label{quant}
\BB_1+\BB_2=\pi N\ ,
\ee
with $N$ a large integer, and
\bb\label{bb1}
\BB_1=\int\ P_v\ dW=\int \Pi_v dv=
\frac{\Sigma}{\sqrt{2\alpe}} E \int_{v_2\rightarrow 0}^{v_c}\ dv\ 
\frac{\sqrt{1-(v/v_c)^3}}{v^2}\ ;
\ee
\bb\label{bb2}
\BB_2=\int_{v_c}^{v2\rightarrow 0}\ \Pi_t\ dt=
\frac{\Sigma}{4\pi\alpe} E \Delta t_{bnd}\ .
\ee
The reader may be concerned with the inclusion of $\BB_2$.
The necessity of this term can be seen from~\pref{prop}; the
exponent in the propagator is the sum of
$\BB_1$ {\em and} $\BB_2$. The physical origin of the quantization condition
is periodicity of this expression, hence the requirement that the
exponent of~\pref{prop} be quantized.
Since our Hamiltonian vanishes, the `counter' of wavefunction nodes
$\oint p\ dx$ is the same as the classical action. $E$ appears in
the quantization condition as a parameter tuning the shape of the
potential of the one dimensional quantum mechanics problem; and the
Bohr-Sommerfeld quantization determines for what potential is
there a `zero energy' state with an integer number of nodes.

The second more important
concern the reader must have has to do with the fact that,
for the bounded motion, we have limits of integration
in~\pref{bb1} and~\pref{bb2} extending to
$v_2\rightarrow 0$. But the latter involves
venturing into forbidden domains of the spacetime if $M$ is taken finite, as 
dictated by~\pref{curvec}. To appreciate the problem at hand
further, let us look at the
behavior of~\pref{bb1} and~\pref{bb2} near $v_2\sim 0$.

For the first term in~\pref{quant}, we have
\bb\label{div1}
\BB_1\sim -\frac{\Sigma}{\sqrt{2 \alpe}} E
\frac{1}{v_2}+\mbox{finite}\ .
\ee
Hence it is divergent as $v_2\rightarrow 0$, since the momentum
canonical to $v$ blows up at the origin (while $\dot{v}$ vanishes).
This divergence arises in the IR of the NCOS theory.

Looking at the second contribution, we get, using~\pref{deltatbnd} 
in~\pref{bb2}
\bb\label{div2}
\BB_2=\frac{\Sigma}{\sqrt{2 \alpe}} E \frac{1}{v_2}+\mbox{finite}\ .
\ee
The origin of this divergence is the infinite redshift
at $v_2\sim 0$ with respect to the time variable $t$; a phenomenon well
known from dynamics near a black hole horizon. And indeed this term cancels
precisely the divergence arising in~\pref{div1}, leading to the 
finite action in~\pref{sclbnd} and~\pref{quant}. The important conclusion
is that the region near $v_2\sim 0$ contributes negligibly to~\pref{quant}
for $v_2\ll 1$ in $\int_0^{v_2} dv$.

For finite $M$,
we then consider taking $M$ large (and $\Sigma/\le\gg 1$ if needed), 
so that the spacetime
is reliable up to very small values in $v$ and the asymptotic expansions
used above are therefore valid. Then, the contribution of the forbidden
region, as we approach it from larger values of $v$, is parametrically
small in the quantization equation~\pref{quant}.

All this still assumes that, 
to leading order in the dynamics of the
closed string, we have bounded motion, with a turning point at $v_c$;
in particular, that our ignorance of the details of the dynamics at
the origin of the radial coordinate $v$ does not change the fact that,
to this level of
semiclassical approximation, the closed string slows down to a virtual
stop as it approaches
the horizon, bounces back, and hence is effectively in
a box. Note that there is no need however of any concern in this regard
for the regime where $M\rightarrow\infty$, as discussed 
in Section~\ref{valsec}. In that strict limiting case, the dynamics
is reliable all the way to $v=0$; and our picture of bounded motion
is correct. Note that no observable,
such as the action or $v_c$, associated
with this dynamics depends on $M$ explicitly; and, hence, the limit 
is regular.

For large but finite $M$,
the regularity of the classical action near $v\sim 0$ gives us some
confidence that, at least as we approach this unknown region,
the dynamics appears well controlled. As we argue in the Discussion section,
we may expect however that
there will be processes involving the splitting and joining the 
closed string; these effects being
subleading in a perturbative expansion in the local value of the
dilaton, and hence involving an expansion in $1/M$.
We then propose a scenario whereby
these effects may be incorporated in our formalism as 
transitions between energy levels of different winding numbers. 
Hence, the spectrum we derive may still be
the leading effect in determining the dynamics. More discussion about this 
may be found in the Discussion section.

Finally,
it is important to emphasize that, despite the fact there
is an infinite redshift at $v\sim 0$, one is able to extract a finite
spectrum for the energy, due to the cancelations detailed earlier. Along
this line of thought, let us also note that we have checked that the phase
factor computed above is additive as the string oscillates between $v=0$
and $v=v_c$ in finite proper time; 
as opposed to perhaps canceling because of some sign flip.

The optimistic reader may wonder whether using the Bohr-Sommerfeld
condition was somewhat too reserved; why not apply the full WKB-like machinery,
in an attempt to estimate the ground state energy of the spectrum near
$N\sim 0$. In fact, expanding the spectrum for small $N$, we do indeed find
some non-zero ground state energy. 
At issue however is a numerical shift on the right hand side
of equation~\pref{quant} $N+\mbox{constant}$, this constant
often called the Maslov index which we have not computed. 
It can be determined by analyzing
caustics of the classical trajectory, a subject that connects
to Catastrophe theory.
While as an Armenian, I readily relate to theories of
catastrophes,
venturing into this analysis is uncalled for in this case, for two reasons. 
First,
the approximation scheme we have adopted (sometimes wrongfully referred
to as WKB) is improved for $N\gg 1$, as is well known. Hence, the estimate
for a ground state energy, with numerical accuracy, is unlikely to be reliable.
The overly optimistic reader may suggest that, this spectrum, being
related to center of mass dynamics, is perhaps `protected' in some
unknown sense and for some unknown reason (perhaps supersymmetry). 
While this is an interesting suggestion,
it is difficult to understand controlled fluctuations without analyzing the
symmetry principle at work, which we have not 
explored in our formalism; and
such quantum fluctuations are otherwise
generically likely to wash out our center of mass
spectrum for small values of $N$. Hence,
the $N\gg 1$ is needed for this purpose 
as well, and the issue of estimating the ground state energy is
deferred to a more complete analysis that studies, amongst other
effects, the role of supersymmetry.

%\bibliography{biblio}
%\bibliographystyle{utphys}

\providecommand{\href}[2]{#2}\begingroup\raggedright\endgroup

\end{document}